%% file: 0.neurips_2022.tex
\documentclass{article}

\PassOptionsToPackage{square, numbers}{natbib}
\usepackage[preprint]{neurips_2022}

\usepackage[utf8]{inputenc} % allow utf-8 input
\usepackage[T1]{fontenc}    % use 8-bit T1 fonts
\usepackage{hyperref}       % hyperlinks
\usepackage{url}            % simple URL typesetting
\usepackage{booktabs}       % professional-quality tables
\usepackage{amsfonts}       % blackboard math symbols
\usepackage{nicefrac}       % compact symbols for 1/2, etc.
\usepackage{microtype}      % microtypography
\usepackage{xcolor}         % colors
\usepackage{wrapfig}

\usepackage[algoruled, vlined, linesnumbered, commentsnumbered]{algorithm2e}
\usepackage{amsmath}
\usepackage{bm}
\usepackage{graphicx}
\usepackage{multirow}

\usepackage{caption}
\usepackage{subcaption}
\usepackage{bbding}
\usepackage{enumitem}
\usepackage{bbm}
\usepackage{color}
\usepackage{colortbl}
\definecolor{Gray}{gray}{0.9}

\usepackage{comment}

\newcommand*{\vpara}[1]{\vspace{0.1in}\noindent\textbf{#1}}

\newcommand*{\Real}{\ensuremath{\mathbb{R}}}
\newcommand*{\mat}[1]{\ensuremath{\bm{#1}}}
\newcommand*{\vect}[1]{\ensuremath{\bm{#1}}}
\newcommand*{\set}[1]{\ensuremath{\mathcal{#1}}}

\newcommand{\model}[0]{SCR}%\xspace}

\title{SCR: Training Graph Neural Networks with Consistency Regularization}

\author{%
  Chenhui~Zhang$^\dagger$\thanks{Equal contribution.}\hskip 0.4em,
  Yufei~He$^\ddagger$\footnotemark[1]\hskip 0.4em,
  Yukuo~Cen$^\dagger$,
  Zhenyu~Hou$^\dagger$,
  Wenzheng~Feng$^\dagger$, \\
  {\bf
    Yuxiao~Dong$^\dagger$,
    Xu~Cheng$^\S$,
    Hongyun~Cai$^\S$,
    Feng~He$^\S$
    and Jie~Tang$^\dagger$
  } \\
  $^\dagger$Department of Computer Science and Technology, Tsinghua University \\
  $^\ddagger$Beijing Institute of Technology \\
  $^\S$Interactive Entertainment Group, Tencent \\
  \texttt{ch-zhang15@mails.tsinghua.edu.cn, yufei.he@bit.edu.cn}
}

\begin{document}

\maketitle

\setcounter{footnote}{0}

\input{0.1.abstract}

\input{1.introduction}
\input{2.related}

\input{3.method}
\input{4.experiments}
\input{5.conclusion}

% \section*{References}

\bibliographystyle{abbrvnat}
\bibliography{reference}

\appendix
\input{6.appendix}

\end{document}

%% file: 0.1.abstract.tex
\begin{abstract}
We present the \model{} framework for enhancing the training of graph neural networks (GNNs) with consistency regularization. 
Regularization is a set of strategies used in Machine Learning to reduce overfitting and improve the generalization ability.
However, it is unclear how to best design the generalization strategies in GNNs, as it works in a semi-supervised setting for graph data. 
The major challenge lies in how to efficiently balance the trade-off between the error from the labeled data and that from the unlabeled data. 
\model{}\footnote{The code is publicly available at:
\url{https://github.com/THUDM/SCR/}.} is a simple yet general framework in which we introduce two strategies of consistency regularization to address the challenge above. 
One is to minimize the disagreements among the perturbed predictions by different versions of a GNN model.
The other is to leverage the Mean Teacher paradigm to estimate a consistency loss between teacher and student models instead of the disagreement of the predictions.
We conducted experiments on three large-scale node classification datasets in the Open Graph Benchmark (OGB). %, which are mostly studied recently.
Experimental results demonstrate that the proposed \model{} framework is a general one that can enhance various GNNs to achieve better performance. 
Finally, \model{} has been the \textbf{top-1} entry on all three OGB leaderboards as of this submission.
%performance on all three OGB datasets as of this writing. 

  %\jie{maybe add more results}
\end{abstract}

%% file: 1.introduction.tex
\section{Introduction}%
\label{sec:introduction}

The graph-structured data, made up of nodes and edges, is a flexible and powerful tool to represent connected objects. 
With the great success of deep learning and neural networks in computer vision, natural language processing, and other fields, {graph neural networks} (GNNs) have also achieved significant performance improvements in graph machine learning tasks~\cite{hamilton2017inductive,velivckovic2017graph,xu2018powerful,gilmer2017neural,maron2018invariant}. 
Many of these GNN models  
become the top-1 performers
%achieved state-of-the-art performance 
on the leaderboard of open datasets such as Open Graph Benchmark~\cite{hu2020open}.

However, training a successful 
%deep learning 
GNN model requires a large amount of labeled data, which is difficult to access in 
%many 
realistic scenarios. 
%Many research papers 
Studies have shown that using unlabeled data 
%with small amount of labeled data 
in training can effectively enhance
%improve 
the model performance. % of the model. 
%Because of the characteristics of graph structure, 
The line of 
research on semi-supervised learning on graphs thus has 
%always 
been a hot 
%research 
topic over decades~\cite{zhu2003semi,zhou2003learning,mei2008general,belkin2006manifold}.
%Some work has proposed to use ``pseudo labeling'' to include more unlabeled nodes in training~\cite{sun2021scalable,zhang2021graph,li2018deeper,yang2021self}, 
Some work has proposed to use ``pseudo labeling'' to exploit unlabeled nodes in the training stage~\cite{sun2021scalable,zhang2021graph,li2018deeper}.
%, which can be regarded as a special case of consistency regularization.
Among them, multi-stage self-training methods have achieved the best performance.
%As the name indicated, the 
Its basic idea is to divide the
training procedure 
%is divided 
into several stages.
%Pseudo labels are assigned to unlabeled nodes according to the predictions of the previous stage to participate in the training.
At the beginning of each stage, the training set is expanded by assigning pseudo labels to unlabeled nodes based on the predictions from the previous stage.
This approach effectively utilizes the information of unlabeled nodes, resulting in better performance. However, the multi-stage approach 
%is less elegant and needs 
requires much more training time than its single-stage counterpart.
%lacks a more general explanation. 

% In this work, we study how consistency regularization can further improve the performance of graph neural networks and provide a general explanation for pseudo labeling. 
In this work, we present a simple and general consistency regularization (\model{}) framework to improve 
the performance of graph neural networks under the semi-supervised setting.
%, as shown in Figure~\ref{fig:framework}. 
%we revisit its role in the training of graph neural networks and examine the performance improvement of graph neural networks under the semi-supervised setting. 
%For consistency regularization, there have been many in-depth studies in the field of deep learning. 
Specifically, we propose
%utilize 
two strategies 
%kinds of methods of 
based on consistency regularization for GNNs. 
The first one, referred to as \model{}, is to 
%One method is called simple consistency regularization (SCR) by 
minimize the disagreement among perturbed predictions 
%The perturbed predictions can be 
%that are obtained 
by 
different versions of a GNN model.
The different versions could be obtained by data augmentation or randomness of the given model. 
For example, GRAND~\cite{feng2020graph} uses random propagation to generate a \textit{graph augmentation}. 
For each graph augmentation, we can generate a version of the predictions by the GNN model. 
By minimizing the disagreement among the predictions from different versions, 
\model{} is able to improve the generalization ability of the GNN model.

The other \model{} strategy is referred to as 
%called 
mean-teacher consistency regularization (\model{}-m), which leverages the teacher-student paradigm. 
For \model{}-m, following Mean Teacher~\cite{tarvainen2017mean}, we guide the training procedure by
%averaging model weights (between the student) 
calculating a consistency loss between the perturbed prediction from the student and the prediction from the teacher 
rather than minimizing the disagreement of prediction results as in \model{}.
%a method that averages model weights instead of label predictions.
Parameters of the teacher model are directly derived from the \textit{Exponential Moving Average} (EMA) weights of the student model without additional back propagation. In each training step, the parameters of the teacher model will be updated by the student model. 
% and explore the role of consistency regularization in the training of graph neural networks. 
% \vpara{Contributions.}

We conducted experiments on three large datasets for the node prediction task from Open Graph Benchmark (OGB)~\cite{hu2020open}, which are mostly studied recently. 
Specifically, we apply the proposed \model{} technique to two recent GNN architectures, namely SAGN~\cite{sun2021scalable} and GAMLP~\cite{zhang2021graph}. 
Experiments on the three datasets clearly show that \model{} can 
%that our methods can 
effectively improve the performance of the base GNNs, in fact outperforming all entries on the three OGB leaderboards. 
In addition, our study also demonstrates that SCR can improve a wide range of (advanced) GNN models,  such as GraphSAGE~\cite{hamilton2017inductive}, SIGN~\cite{yu2020scalable}, ClusterGCN~\cite{ChiangLSLBH19}, and GraphSAINT~\cite{zeng2019graphsaint}, demonstrating the benefit of \model{} as a general framework for enhancing GNNs.

%by 0.12\%$\sim$0.41\%. 
To sum up, this work makes the following contribution: 
%We sum up the contributions of the paper:
\begin{itemize}[leftmargin=*,itemsep=0pt,parsep=0.2em,topsep=0.3em,partopsep=0.3em]
  \item \textit{Effectiveness}. %Consistency regularization is a significant technique for training of graph neural networks, especially in the semi-supervised manner. The e
  Experimental results demonstrate that the proposed \model{} achieves 
    %\textbf{top-1 performance} 
  the best performance on all the three large datasets from Open Graph Benchmark (OGB). 
  \item \textit{Scalability}. \model{} is a simple and efficient framework with scalable methods of consistency regularization for GNNs, which allows it to scale to graphs with 100 million nodes and 1 billion edges. 
    % , ClusterGCN, GraphSAINT and SIGN
  \item \textit{Flexibility}. \model{}, as a general framework, is very flexible to all genres of graph neural networks, such as SAGN, GAMLP, GraphSAGE, SIGN, ClusterGCN, and GraphSAINT. 
    %Experimental results demonstrate that our methods achieve top-1 results on the three OGB datasets. 
\end{itemize}

% The method of pseudo labeling can be regarded as a special case of consistency regularization. It can be viewed as existing a hidden teacher model in the training process, whose parameters directly copy the model parameters of the previous stage at the beginning of the new stage. Within a single stage, its parameters do not change at all. 

% The whole work is still in the process of improvement. 

% Better performance is to be expected.

% The effectiveness of GNNs in SSL is primarily due to the fact that GNNs can produce supervision signals on unlabeled nodes by propagating the information on labeled nodes along edges.

%Exponential Moving Average (EMA) is proposed in Temporal Ensembling.The basic assumption is that the weight of the model will wobble at the actual best point in the last n steps, so we take the average of the last N steps to make the model more robust.

%% file: 2.related.tex
\section{Related Work}%
\label{sec:related}

%\subsection
\vpara{Graph Neural Networks.}
Due to the great success of deep learning in various areas, much effort has been devoted to generalizing neural networks to graph-structured data, resulting in 
the development of graph neural networks (GNNs).
%CNN is transferred to graph-structured data and is called graph convolutional neural network (GCN), which leads the development of graph neural network (GNN).
% The current GNN model 
%GNN models are mainly divided into two categories: spectral domain based models and spatial domain based models.From a spectral perspective, ~\citet{bruna2013spectral} firstly used the Fourier basis of graphs in the spectral domain to develop graph convolutions.
As a pioneer work, ~\citet{kipf2016semi} proposed graph convolution network (GCN) which adopted Chebyshev polynomials to approximate the graph spectral filter.
After that, spatial domain based GNNs~\cite{hamilton2017inductive,velivckovic2017graph,xu2018powerful} are proposed by generalizing the graph convolution to various neighborhood aggregation functions, making GNNs become the mainstream for graph modeling.
%information to update each node's own representation. 
%This simple and effective approach now becomes mainstream. 
%There are quite a few works that try to go beyond this structure, e.g., GMNN~\cite{qu2019gmnn} and GraphNAS~\cite{gao2019graphnas}.
%Some sampling-based work has also been achieved in the training of scalable graph neural networks,
Based on the formulation of spatial GNNs, there have been several sampling strategies proposed to scale up GNNs on large graphs, e.g., FastGCN~\cite{chen2018fastgcn}, AS-GCN~\cite{huang2018adaptive}, and GraphSAINT~\cite{zeng2019graphsaint}. 
% Several works have tried different approaches to training deeper GNNs~\cite{rahimi2018semi,xu2018representation,armeni2017joint}. 
% The main problem to be solved is over-smoothing~\cite{li2018deeper,klicpera2018predict,zhao2019pairnorm}.

%In the initial GNN models, \textit{feature transformation} and \textit{feature aggregation} are coupled together.

% and achieved state-of-the-art performance on OGB datasets.

% \vpara{Consistency Regularization on Graphs.}
\vpara{Regularization Methods for GNNs.} 
In the meanwhile of developing GNN architectures, several works also focus on improving the training of GNNs with the consistency regularization technique.
%The focus of semi-supervised learning is on how to utilize unlabeled data to improve performance.
%Recent efforts for this problem mainly use data augmentation and consistency regularization to achieve it.
Consistency regularization was first proposed by ~\cite{bachman2014learning}, which employs a consistency loss to enforce the model to give similar predictions among different augmentations of unlabeled data.
%later referred as ``$\Pi$-Model''~\cite{rasmus2015semi}.
% Based on this, subsequent work has improved the way labels are generated.
%Several works further improved this paradigm by developing different data augmentation strategies, e.g., Temporal Ensembling~\cite{laine2016temporal}, Mean Teacher~\cite{tarvainen2017mean}, virtual adversarial training~\cite{miyato2018virtual}, MixMatch~\cite{berthelot2019mixmatch}, and UDA~\cite{xie2019unsupervised}.
%One of the key challenges in consistency regularization is how to improve the quality of the generated labels.
%Temporal Ensembling~\cite{laine2016temporal} uses previous model checkpoints while Mean Teacher~\cite{tarvainen2017mean} uses an exponential moving average of model parameters. UDA~\cite{xie2019unsupervised} and ReMixMatch~\cite{berthelot2019remixmatch} sharpen the soft label to make the model to produce high-confidence predictions. UDA further reinforces consistency only when the highest probability of the predicted category distribution for soft labels is above a threshold. 
%These techniques have achieved great success in semi-supervised image classification, and further motivated researchers to apply them onto GNNs for semi-supervised learning on graphs.
%Some works have explored regularization training methods for GNNs. 
To extend this idea onto graphs, the main efforts concentrates on designing data augmentation strategies for 
For example, 
%such as VBAT~\cite{deng2019batch}, GraphVAT~\cite{feng2019graph}, GraphMix~\cite{verma2019graphmix}, DropEdge~\cite{rong2019dropedge}, NodeAug~\cite{wang2020nodeaug} and GRAND~\cite{feng2020graph}. 
VBAT~\cite{deng2019batch} and GraphVAT~\cite{feng2019graph} utilized virtual adversarial training to generate graph data augmentations. % regularization training on GNNs. 
GraphMix~\cite{verma2019graphmix} borrowed the idea from MixMatch~\cite{berthelot2019mixmatch} by adopting MixUp~\cite{zhang2017mixup} to facilitate GNN training. 
%DropEdge~\cite{rong2019dropedge} applied dropout~\cite{SrivastavaHKSS14} onto graph structure. 
GRAND~\cite{feng2020graph} and NodeAug~\cite{wang2020nodeaug} further explored more complex augmentation strategies for graph data and achieved significant performance gains on small graphs.
%designed graph data augmentation strategies for leveraging consistency regularization to optimize the prediction consistency of unlabeled nodes, and achieved significant performance gains on small graphs. After that, several works~\cite{wang2020graphcrop,hassani2020contrastive,zhao2020data} were proposed to improve GNN performance by designing more advanced data augmentation strategies for graph structure.
However, these data augmentation methods are usually time-consuming especially on large-scale graphs because it needs to be executed in each training step. 

A similar technique to consistency regularization is self-training, which employs the ``pseudo-labels'' of unlabeled nodes to facilitate model training.  
%\subsection
%\vpara{Self-training Methods for GNNs.}
%Using ``pseudo-label'' to improve GNN performance becomes a common practice for semi-supervised learning tasks. 
\citet{li2018deeper} was an early-stage work to explore co-training and self-training of GCNs, which significantly boosted the performance of GCNs when using very few labels in training. 
\citet{sun2020multi} proposed a {Multi-Stage Self-Supervised} (M3S) training algorithm, which used clustering approach to construct pseudo-labels. % and then utilized the aligning mechanism. 

%% file: 3.method.tex
% \begin{figure*}
%     \centering
%     \includegraphics[width=0.85\textwidth]{figures/crgnn-v2.pdf}
%     \caption{This diagram shows the our proposed \model framework. The total training loss is divided into supervised loss and consistency loss. The right part shows how to estimate pseudo labels. Specific to the \model-m, for the unsupervised part, the teacher model and student model adopt the same input to calculate consistency loss between their outputs. During the training process, the weights of student model are updated by gradient descent, while the weights of teacher model is updated by EMA of student model. For \model-1, the same data is fed into the same model twice to compute the consistency loss without using the teacher model.}
%     \label{fig:my_label}
% \end{figure*}

% \begin{figure}
%     \centering
%     \includegraphics[width=0.42\textwidth]{figures/crgnn_figure-xxv.pdf}
%     \caption{Caption}
%     \label{fig:my_label2}
% \end{figure}

% The left part shows how decoupled GNN performs propagation of features and labels to obtain preprocessing embeddings. The right part shows the training process of MCR (\textit{Mean-Teacher Consistency Regularization}) and SCR (\textit{Simple Consistency Regularization}) approaches. Labeled data is fed into the student model to compute the supervised loss between the output prediction and one-hot label.

% \begin{figure}
%     \centering
%     \includegraphics[width=0.25\textwidth]{figures/crgnn-v3.pdf}
%     \caption{Option 2}
%     \label{fig:my_label}
% \end{figure}

\section{Preliminaries}%
\label{subsec:prelims}

%In this paper, w
%We use bold uppercase letters (e.g., $\mat{A}$) for matrices, and bold lowercase letters (e.g., $\vect{x}$) for vectors.
% The transpose of a matrix $\mat{A}$ (resp. a vector $\vect{x}$) is denoted by $\mat{A}^{\intercal}$ (resp. $\vect{x}^{\intercal}$).
% For any matrix $\mat{A}$, we use $\mat{A} [i, j]$, $\mat{A} [i, :]$, and $\mat{A} [:, j]$ to represent its $(i, j)$-th entry, $i$-th row and $j$-th column, respectively.
% Likewise, we use $\vect{x} [i]$ to denote the $i$-th entry of $\vect{x}$.
We denote a graph with $N$ nodes by $G = (\set{V}, \set{E})$ where $\mathcal{V}$ is its node set, and $\set{E} \subseteq \set{V} \times \set{V}$ is a set of edges.
The adjacency matrix of $G$ is denoted by $\mat{A} \in \Real^{N \times N}$ with its $(i, j)$-th entry $\mat{A} [i, j] = \mathbbm{1} ((i, j) \in \set{E})$ indicating whether there is an edge from node $i$ to node $j$.
We also assume that each node $i \in \set{V}$ is associated with a feature vector $\vect{x}_i \in \Real^{d}$.% where $d$ is the number of features.

% Semi-supervised learning provides an efficient way to improve the performance of deep neural networks with abundant unlabeled data.
We illustrate our approach in the context of \emph{node classification}, though it can be easily generalized to other tasks such as link prediction and subgraph classification.
In the setting of node classification, we are given a graph $G = (\set{V}, \set{E})$ and a set of labeled nodes, denoted by $\set{V}_L \subset \set{V}$.
Each labeled node $i \in \set{V}_L$ is associated with a one-hot vector $\vect{y}_i \in \{ 0, 1 \}^{C}$ which encodes its ground-truth class.
$C$ is the number of predefined classes.
Our goal is to learn a function $f_{\theta} (i \mid G)$ parameterized by $\theta$ which can predict the correct class for a given unlabeled node $i$.

\vpara{Decoupled GNNs}. 
Conventional GNN architectures alternate between the \textit{feature transformation} and \textit{feature aggregation} operators.
E.g., in the GCN model~\cite{kipf2016semi}, A feature transformation operation and a feature aggregation operator are coupled together to form a graph convolution layer.
%But in fact,
However, recent studies also suggest that this coupling paradigm is practically unnecessary and develop a series of decoupled GNNs~\cite{liu2020towards,wu2019simplifying,he2020lightgcn}. Such GNNs first propagate the input feature $\mat{X}$ on graph for $K$ steps by iteratively calculating:
\begin{equation}
\mat{X}^{(k)} = \hat{\mat{A}}^k \mat{X}^{(k-1)}, \forall k = 0, 1, \dots, K
\end{equation}
where $\mat{X}^{(0)} = \mat{X}$ and $\hat{\mat{A}} \in \Real^{N  \times N}$ is an aggregation matrices. 
The generated feature matrices $\{\mat{X}^{(0)}, \mat{X}^{(1)}, \cdots, \mat{X}^{(K)}\}$ are then fed into a neural network $f_\theta$ to make predictions:
\begin{equation}
    \hat{\mat{Y}} = f_{\theta}(\{\mat{X}^{(0)}, \mat{X}^{(1)}, \cdots, \mat{X}^{(K)}\}),
\end{equation}
where $\hat{\mat{Y}}\in [0,1]^{N\times C}$ denotes the generated prediction probability matrix. And the model is usually trained with the cross entropy loss on labeled nodes:
\begin{equation*}
  \mathcal{L}_{sup} = \frac{1}{|\set{V}_L|} \sum_{i \in \set{V}_L} \mathtt{CrossEntropy} (\vect{y}_i, \vect{\hat{y}}_i).
\end{equation*}
%The above loss function indicates that GNNs are only trained on labeled nodes.
In practice,
%al scenarios, 
the labeled nodes can be scarce while there are abundant unlabeled nodes in the graph.
In this paper, we will investigate how much performance can be gained when training GNNs with both labeled and unlabeled nodes. To this end, we adopt two advanced decoupled 
%developed 
GNN architectures---SAGN~\cite{sun2021scalable} and GAMLP~\cite{zhang2021graph} as backbone models due to their SOTA performances on quite a few open datasets.

\begin{figure*}[t]
    \centering
    \includegraphics[width=0.99\textwidth]{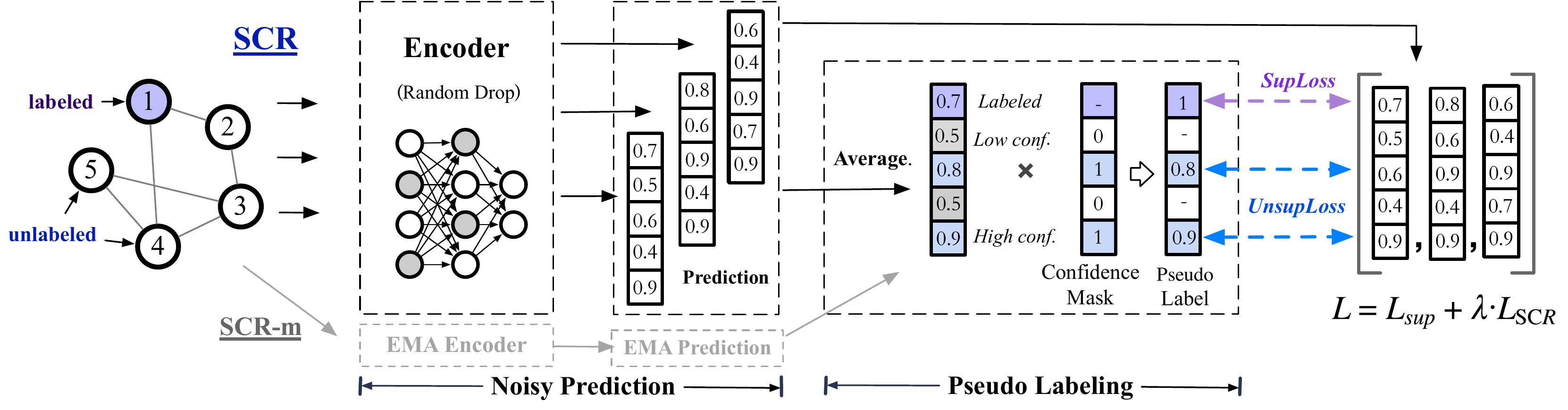}
    \caption{Illustration of the proposed \model{} framework. 
    \textmd{
    The total training loss is divided into the supervised loss and unsupervised consistency loss. 
    The difference between \model{} and \model{}-m mainly lies on how to estimate pseudo labels. 
    %For \model{}, the same input data is fed into the same model with dropout several times to compute the perturbed predictions. 
    In \model{}, the pseudo label is the average of the outputs over multiple runs, while in \model{}-m, the pseudo label is predicted by an EMA teacher.
    Confidence-based masking is used to get rid of pseudo labels with high uncertainty.
    % The predictions will be averaged and become pseudo labels through the confidence mask. 
    % The consistency loss will be estimated between pseudo labels and several versions of model predictions. 
    % Specific to the \model{}-m, we use another EMA encoder as a teacher model. 
    % The teacher model and student model adopt the same input to calculate consistency loss between their outputs. During the training process, the weights of student model are updated by gradient descent, while the weights of teacher model are updated by EMA of the student model.
    %}
    }
    }
    \label{fig:framework}
\end{figure*}

%\subsection{The SCR Framework}
\section{The \model{} Framework}%
\label{sec:method}

In this section, we introduce the proposed \model{} framework to %can help
%an effective regularizer for 
improve the training of graph neural networks.
\model{} is based on the concept of \emph{consistency regularization}~\cite{Sajjadi:NeurIPS16,xie2019unsupervised}.
%, a widely used technique in semi-supervised learning (SSL).
Briefly, consistency regularization encourages the learned model to be in line with the \emph{low-density separation assumption}~\cite{chapelle2006semi}, i.e., the learned decision boundary should lie in the low-density region.
In other words, a small perturbation on the input data should not change the prediction significantly.
% Following this idea, a straightforward implementation of the consistency regularization is to minimize the distance among the outputs generated by different corrupted versions of the input.

%Figure~\ref{fig:framework} illustrates one example of the proposed \model framework.

Figure~\ref{fig:framework} illustrates the process of training a graph neural network under the \model{} framework. 
%It requires a stochastic encodes to can produce different outputs for the same input. 
%This can be easily fulfilled by adding dropout regularization~\cite{SrivastavaHKSS14} to the model.
Given an input graph with 1 labeled node and 4 unlabeled nodes on the left, it is fed into an encoder $S$ times (e.g., $S = 3$ in the figure) to get $S$ \textbf{noisy predictions} (Cf. \S~\ref{sec:noisy_pred}).
These predictions are different due to the stochastic property of the encoder, e.g., a neural network with dropout. 
Then, we assign a \textbf{pseudo label}, e.g., the average of these noisy predictions, to every unlabeled node (Cf. \S~\ref{sec:pseudo_label}).
Next, a confidence mask is computed according to the pseudo labels, which filters out unlabeled nodes that the model cannot predict well.
The \textbf{loss function} consists of two components.
The first one is the standard cross entropy loss between ground-truth labels and predictions from labeled nodes. The second component is evaluated based on the confident unlabeled nodes, and used to penalize a distance metric between predictions and pseudo labels (Cf. \S~\ref{sec:loss}). 

The \model{} framework is a general consistency regularization technique for improving graph learning models. 
For example, the encoder can be a simple MLP or any graph neural networks, e.g., GCN~\cite{kipf2016semi}, GraphSAGE~\cite{hamilton2017inductive}, and GAMLP~\cite{zhang2021graph}. 
% \dong{to chenhui: add }

\begin{comment}
Algorithm~\ref{algo:scr} gives an overview on how we train a GNN model $f_{\theta} (i \mid G)$ under the the SCR framework.
At each training step, we are given a batch of labeled nodes and unlabeled nodes.
We first generate a total of $S$ \textbf{noisy predictions} for every node in the batch, which will be detailed in the following sections.
We also assign a \textbf{pseudo label} for every unlabeled node.
The loss function consists of two components.
The first one
% component 
is the standard cross entropy loss between ground-truth labels and predictions from labeled nodes, while the second component, evaluated on unlabeled nodes, penalizes a distance metric between predictions and pseudo labels.
Figure~\ref{fig:framework} also illustrates this process.
\end{comment}

\subsection{Noisy Prediction Generation}
\label{sec:noisy_pred}

Injecting noise to the model in the training stage is a simple yet efficient technique to regularize deep neural networks.
The noise can be added into the input, e.g., data augmentation~\cite{xie2019unsupervised}, or directly added into the model, e.g., dropout~\cite{SrivastavaHKSS14}. 
For example, there are many semantic-preserving transformations, such as flipping, rotation and color jitter, to augment image data.
%Thus, d
%Data augmentation is widely used in consistency regularization algorithms for image classification~\cite{xie2019unsupervised}.
%However, these operations have no direct analogs 
However, these cannot be directly applied to the graph data. 
Furthermore, existing data augmentation approaches for graphs, such as DropNode~\cite{feng2020graph} and DropEdge~\cite{rong2019dropedge}, are operated on the entire graph. 
This means that we need to load the entire graph into the memory at each training step, which is time-consuming and memory inefficient for large graphs.

In \model{}, we only use the simple dropout strategy to generate noisy predictions.
%, making it 
For each node $i \in \set{V}$, we can obtain a total of $S$ noisy predictions $\{ \hat{\vect{y}}_i^{(s)} \}_{s = 1}^S$ by evaluating the model $f_{\theta} (i\mid G)$ for $S$ times under different dropout conditions.
During the training stage, the dropout regularization randomly disables a predefined proportion of hidden units of the model to avoid co-adapting and overfitting, and the output is a random variable.
Therefore, multiple evaluations of the same input will yield different results.
According to the low-density separation assumption~\cite{chapelle2006semi}, this difference should be small, given that the input data is the same. 
Thus, minimizing the output difference caused by the dropout regularization is reasonable. 
Moreover, by using dropout, the \model{} framework is in nature a lightweight and scalable method that can regulate models on large-scale graphs. 

%Compared to other consistency regularization methods, the SCR framework does not rely on the data augmentation.
%Instead, we only exploit the inherent randomness in the model itself, i.e., the dropout regularization~\cite{SrivastavaHKSS14}.
%Dropout is a powerful regularization technique to increase the generalization ability of the model.

\begin{algorithm}[t]
  \SetKwInOut{Input}{Input}
  \SetKwInOut{Output}{output}
  \Input{Graph $G = (\set{V}, \set{E})$; Labeled nodes $\set{V}_L$; Ground-truth for labeled nodes $\{ \vect{y}_i \mid i \in \set{V}_L \}$; Number of views $S$; Number of Training steps $N_t$; Number of labeled nodes in a batch $N_L$; Number of unlabeled nodes in a batch $N_U$.}
  \For{$t \leftarrow 1$ \KwTo $N_t$}{
    Sample $N_L$ labeled nodes $\set{B}_L \subseteq \set{V}_L$ and $N_U$ unlabeled nodes $\set{B}_U \subseteq \set{V} \setminus \set{V}_L$\;  
    \For{$i \in \set{B}_L \cup \set{B}_U$}{
      \For{$s \leftarrow 1$ \KwTo $S$}{
        $\hat{\vect{y}}_i^{(s)} \leftarrow f_{\theta} (i \mid G)$ with dropout\;
      }
     }
    \For{$i \in \set{B}_U$}{
      Generate pseudo label $\bar{\vect{y}}_i$ using Equation~\ref{eq:scr_1} or~\ref{eq:scr_m}\;
      $\bar{\vect{y}}_i^{(sharp)} \leftarrow \mathtt{sharpen} (\bar{\vect{y}}_i)$\;
    }
    Compute the loss $\mathcal{L}$ using Equation~\ref{eq:loss}\;
    Update parameters $\theta$ by minimizing $\mathcal{L}$\;
  }
  \caption{\label{algo:scr}Pseudocode for the SCR framework.}
\end{algorithm}

\subsection{Pseudo Labeling}
\label{sec:pseudo_label}

\begin{comment}
A simple form of consistency regularization is to minimize the average distances between any pair of the generated noisy predictions for every node $i$:
\begin{equation}
  \label{eq:con_reg}
  \mathcal{L}_{CR} = \frac{1}{|\set{V}_U|} \sum_{i \in \set{V}_U}\frac{2}{S (S - 1)} \sum_{1 \leq s < s' \leq S} \mathtt{dist} \left( \hat{\vect{y}}_i^{(s)}, \hat{\vect{y}}_i^{(s')} \right),
\end{equation}
where $\set{V}_U = \set{V} \setminus \set{V}_L$ is the set of unlabeled nodes, and $\mathtt{dist}$ is a function measures the distance between two predictions.
The popular distance measures include the mean squared error, the Kullback-Leibler divergence and the Jensen-Shannon divergence.
Notice that when $S = 2$, Equation~\ref{eq:con_reg} is the same as the unsupervised loss term used in the $\Pi$-model~\cite{laine2016temporal}.

The problem with the consistency regularizer defined by Equation~\ref{eq:con_reg} is that the predictions $\hat{\vect{y}}_i^{(s)}$ can be of high variance and low confidence due to the dropout noise.
This will lead to an instability during training and reduce the effectiveness of the consistency term.

To relief this,
\end{comment}

In order to extract information from the unlabeled data, we need to construct learning tasks on them.
\emph{Pseudo labelling} is a popular method which produces pseudo labels for unlabeled data using the model itself and trains the model against them.
%\emph{pseudo label} 
% of node $i$ for every unlabeled node $i$, and enforce all the noisy predictions of $i$ to be close to its pseudo label.
Denoting the pseudo label of node $i$ as $\bar{\vect{y}}_i$, the regularization term used in \model{} framework is defined as:
\begin{equation}
  \label{eq:scr}
  \mathcal{L}_{SCR} = \frac{1}{S \cdot |\set{V}_U|} \sum_{i \in \set{V}_U} \sum_{s = 1}^S \mathtt{dist} \left( \bar{\vect{y}}_i, \hat{\vect{y}}_i^{(s)} \right).
\end{equation}
where $\set{V}_U = \set{V} \setminus \set{V}_L$ is the set of unlabeled nodes, and $\mathtt{dist}$ is a function measures the distance between two predictions, such as the mean squared error and the Kullback-Leibler divergence.

We describe two strategies to generate the pseudo labels --- \model{} and \model{}-m.
In \model{}, the pseudo label for node $i$ is defined as the average of its noisy predictions, i.e.,
\begin{equation}
  \label{eq:scr_1}
  \bar{\vect{y}}_i = \frac{1}{S} \sum_{s = 1}^S \hat{\vect{y}}_i^{(s)}.
\end{equation}
In \model{}-m, we use a teacher network to generate pseudo labels:%, i.e.,
\begin{equation}
  \label{eq:scr_m}
  \bar{\vect{y}}_i = f_{\theta'} (i \mid G),
\end{equation}
where $f_{\theta'} (i \mid G)$ is the teacher network.
Motivated by the Mean-Teacher paradigm described in~\cite{tarvainen2017mean}, the teacher network shares the same network architecture with the student model $f_{\theta} (i \mid G)$, and its parameters at training step $t$, expressed by $\theta_t' = \alpha \theta_{t - 1}' + (1 - \alpha) \theta_t$, are an exponential moving average of the parameters of the student where $\alpha \in (0, 1)$ is a decay rate.
%More specifically, given a decay rate $\alpha \in (0, 1)$, the parameters $\theta'$ is updated at the end of the training step as follows:
%\begin{equation*}
%  \theta' = \alpha \theta' + (1 - \alpha) \theta.
%\end{equation*}
%The teacher network can be treated as a temporal ensemble of the original model, and is expected to produce more accurate and stable predictions.

\vpara{Pseudo label sharpening.}
After obtaining the pseudo label via Equation~\ref{eq:scr_1} or~\ref{eq:scr_m}, we further apply a sharpening function to reduce the entropy of the pseudo label.
\begin{equation*}
    \bar{\vect{y}}^{(sharp)} [c] = \mathtt{sharpen} (\bar{\vect{y}}) [c] = \frac{(\bar{\vect{y}} [c])^{1 / T}}{\sum_{c' = 1}^C (\bar{\vect{y}} [c'])^{1 / T}}
\end{equation*}
where $T \in (0, 1)$ is a temperature hyperparameter controlling the sharpness of the pseudo label.
This operation is motivated by the idea of \emph{entropy minimization}~\cite{GrandvaletB04} which encourages the model to produce low-entropy (high-confident) predictions, and is widely used in other consistency regularization methods~\cite{xie2019unsupervised,berthelot2019mixmatch,feng2019graph}.

\subsection{Loss Function}
\label{sec:loss}

%at the beginning of this section, 
As aforementioned, the loss function of the \model{} framework consists of two parts.
The first part is evaluated on the labeled nodes while the second part is evaluated on the unlabeled nodes.
At each training step, we randomly select a batch of nodes which is a mixture of labeled and unlabeled nodes.
We use $\set{B}_L \subseteq \set{V}_L$ and $\set{B}_{U} \subseteq \set{V}_U$ to denote the sets of labeled nodes and unlabeled nodes in the batch.
%The labeled nodes are processed by the student to build the supervised loss, while the unlabeled nodes are processed by both the teacher network and the student to build the unsupervised loss.
The loss function of the batch can be expressed as:
\begin{equation}
  \label{eq:loss}
  \mathcal{L} = \frac{1}{S \cdot N_L} \sum_{i \in \set{B}_L} \sum_{s = 1}^S \mathtt{CrossEntropy} (\vect{y}_i, \hat{\vect{y}}_i^{(s)}) + \frac{\lambda}{S \cdot N_U} \sum_{i \in \set{B}_U} \sum_{s = 1}^S \mathtt{dist} \left( \bar{\vect{y}}_i^{(sharp)}, \hat{\vect{y}}_i^{(s)} \right)
\end{equation}
where $N_L$ and $N_B$ denote the size of $\set{B}_L$ and $\set{B}_U$, respectively.
$\lambda$ is a hyperparameter that controls the strength of the regularizer.
Following~\cite{tarvainen2017mean,xie2019unsupervised,berthelot2019mixmatch}, we do not propagate gradients through pseudo labels.

\begin{comment}
The parameters of the student are updated with the gradient descent algorithm:
\begin{equation*}
  \vect{\theta}_s = \vect{\theta}_s - \gamma \frac{\partial \mathcal{L}}{\partial \vect{\theta}_s},
\end{equation*}
where $\gamma$ is the learning rate.

\vpara{Inference stage.}
When predicting the label for a given node $i$, only the student is used to infer the class distribution.
The final label of node $i$, denoted as $\hat{y}_i$, is defined as the class with the highest probability in the distribution $p_{Y \mid i, \vect{\theta}_s}$:
\begin{equation*}
 \hat{y}_i = \operatorname{argmax}_{c \in \{1, \dots, C\}} p_{Y \mid i, \vect{\theta}_s} (c).
\end{equation*}
\end{comment}

\subsection{Training Algorithm}

Algorithm~\ref{algo:scr} summaries the training process of the \model{} framework.
In this section, we present several useful techniques that help improve the training effectiveness.
%performance of the MCR.
These techniques can be easily extended to other consistency regularization frameworks.

\vpara{Confidence-based masking.}
Minimizing the regularization term via Equation~\ref{eq:scr} is to pull the predictions of the model towards the generated pseudo labels.
It is desirable when the pseudo labels are of high confidence.
However, the pseudo labels are not always reliable, especially at the beginning of the training process.
Poor pseudo labels could even hinder the model from fitting to the labeled data.
To solve this, we filter out low-quality predictions with the confidence-based masking strategy~\cite{xie2019unsupervised}.
Specifically, we maintain a confident node set $\set{V}_C = \{ i : i \in \set{V}_U \cap \max_{c\in \{1, \dots, C\}} \bar{\vect{y}} [c] \geq \eta \}$ in the training stage whose elements are unlabeled node with highly skewed predictions.
Here, $\eta \in [0, 1]$ is a threshold that controls the size of the confident set.
This is based on the assumption that the skewed predictions are more likely to be a confident one as they are far away from the decision boundary.
At each training step, the unlabeled nodes in the minibatch are selected from this confident node set $V_C$ instead of the entire unlabeled node set $V_U$.
In our experiments, the confident node set is updated every $\beta$ epochs.
Furthermore, we find it helpful to gradually lower the threshold $\eta$ during the training process in some situations, which allows more nodes to contribute to the consistency regularization as the training proceeds.

\vpara{Warmup for \model{}-m.}
When training \model{}-m, we found that
%find it would be very helpful with a warmup strategy. 
the teacher network performs worse than the student network at the early stage of the training, and thus, 
%can hardly bring positive effects to the training of the model.
%It 
may 
%even 
hurt the stability of the student network.
%This may be due to the difference between their parameter updating rules.
This might be due to that
the student network is updated via gradient descent, and fits to the labeled data very quickly, while the teacher network is updated via EMA, and lags behind the student at the early stage of training.
% which makes the teacher network performs worse than the student network.
% thus it is suggested that the teacher network should use low EMA decay $\alpha$ at the beginning of training to forget the old, inaccurate, student weights quickly~\cite{tarvainen2017mean}.
% However, our experimental experience showed that, in the initial training epochs, an inadequate teacher can hardly bring positive effects to the training of the student, sometimes even reduce the stability of the student.
%It is not beneficial to use teacher model to guide student's training at the beginning of training.
%So, 
%Therefore, 
Accordingly, we design a warmup stage for \model{}-m by disabling its regularizer at the first $\tau$ training epochs, preventing the model from fitting to the predictions made by the insufficiently-trained teacher model.

%% file: 4.experiments.tex
\section{Experiments}%
\label{sec:experiments}

% \end{wraptable}

\begin{comment}
\begin{table}[tb]
\centering
\small
\caption{\label{tab:dataset}Statistics of three OGB datasets.}
%   \vspace{-0.1in}
\begin{tabular}{lrrrr}
\toprule
Datasets & \#Nodes & \#Feat. & \#Edges & \#Classes \\
\midrule 
% Cora  & 2,708 & 1,433 & 5,429 & 140/500/1000 & 7 \\
% Citeseer &3,327 & 3,703 & 4,732 & 120/500/1000 & 6   \\
% Pubmed  & 19,717 & 500 & 44,338 & 60/500/1000 & 3   \\
ogbn-products &2,449,029 & 100 & 61,859,140  & 47 \\
ogbn-mag & 1,939,743& 128 &21,111,007 &349  \\
ogbn-papers100m & 111,059,956& 128 &1,615,685,872  &172\\
\bottomrule
\end{tabular}
% \vspace{-3mm}
\end{table}
\end{comment}

% \begin{table*}[thbp]
% \centering
% \caption{\label{tab:dataset}Statistics of datasets}
% \begin{tabular}{l|rrrrr}
% \toprule
% Datasets & \#Nodes & \#Features & \#Edges & Train/Valid/Test Nodes & \#Classes \\
% \midrule 
% % Cora  & 2,708 & 1,433 & 5,429 & 140/500/1000 & 7 \\
% % Citeseer &3,327 & 3,703 & 4,732 & 120/500/1000 & 6   \\
% % Pubmed  & 19,717 & 500 & 44,338 & 60/500/1000 & 3   \\
% ogbn-products &2,449,029 & 100 & 61,859,140 &196k/49k/2204k & 47 \\
% ogbn-papers100m & 111,059,956& 128 &1,615,685,872 &1207k/125k/214k &172\\
% ogbn-mag & 1,939,743& 128 &21,111,007&626k/66k/37k &349  \\
% \bottomrule
% \end{tabular}
% \end{table*}

%\subsection{Comparison against SOTA}

%In this section, we conduct experiments on 
We evaluate the \model{} (and \model{}-m) on
the node classification task, which is one of the most popular benchmark tasks for graph learning. 
We use three large datasets from Open Graph Benchmark (OGB)~\cite{hu2020open},
% which is proposed to provide high-quality datasets with different scales and domains, as well as the evaluation. 
% To demonstrate the effectiveness of our methods, we conducted experiments on three OGB datasets for node-level prediction,
including ogbn-products, ogbn-mag and ogbn-papers100M.
Table~\ref{tab:dataset} summaries statistics of the three benchmark datasets
We follow exactly the experimental procedure suggested by OGB, such as features, data splits and evaluation protocol.
More details are covered in Appendix.% ~\ref{sec:appendix}.

\begin{table}[tb]
\centering
\small
\caption{\label{tab:dataset}Statistics of three OGB datasets.}
%   \vspace{-0.1in}
\begin{tabular}{lrrrr}
\toprule
Datasets & \#Nodes & \#Feat. & \#Edges & \#Classes \\
\midrule 
% Cora  & 2,708 & 1,433 & 5,429 & 140/500/1000 & 7 \\
% Citeseer &3,327 & 3,703 & 4,732 & 120/500/1000 & 6   \\
% Pubmed  & 19,717 & 500 & 44,338 & 60/500/1000 & 3   \\
ogbn-products &2,449,029 & 100 & 61,859,140  & 47 \\
ogbn-mag & 1,939,743& 128 &21,111,007 &349  \\
ogbn-papers100m & 111,059,956& 128 &1,615,685,872  &172\\
\bottomrule
\end{tabular}
% \vspace{-3mm}
\end{table}

% Table~\ref{tab:dataset} shows the dataset statistics.

%As suggested by OGB, we repeat each experiment for 10 times with random seeds and report the average and the standard deviation of the classification accuracy.  
%On ogbn-papers100M, the reported results are based on 3 independent runs. 
%The results of other baselines are directly taken from the OGB leaderboards.
% MOVED TO APPENDIX

% We report the results on the OGB leaderboards for baselines. 
Overall, the proposed \model{} framework has been the \textbf{top-1 entry} on all the three OGB datasets as of today---the submission deadline.

\subsection{Comparison with Peer Methods}

\begin{table}[tb]
% \begin{wraptable}{r}{10cm}
  \centering
  \caption{Accuracy on ogbn-products. Results with gray are obtained by our proposed framework.}
   \label{tab:products}
  \begin{tabular}{llccc}
    \toprule
    Methods & Arch. & C\&S & Validation & Test \\
    \midrule
    - & MLP & - & 75.54\tiny{$\pm$0.14} & 61.06\tiny{$\pm$0.08} \\
    - & MLP & \Checkmark & 91.47\tiny{$\pm$0.09} & 84.18\tiny{$\pm$0.07} \\
    - & GCN & - & 92.00\tiny{$\pm$ 0.03} & 75.64 \tiny{$\pm$ 0.21} \\
    - & GraphSAGE & - & 92.24\tiny{$\pm$0.07} & 78.50\tiny{$\pm$0.14} \\
    - & SIGN & - & 92.99\tiny{$\pm$0.04} & 80.52\tiny{$\pm$0.16} \\
    - & SAGN & - & 93.09\tiny{$\pm$0.04} & 81.20\tiny{$\pm$0.07} \\
    - & GAMLP & - & 93.12\tiny{$\pm$0.03} & 83.54\tiny{$\pm$0.09} \\
    \midrule
    % SLE & SAGN & - & 93.09\tiny{$\pm$0.07} & 84.68\tiny{$\pm$0.12} \\
    % SLE & SAGN & \Checkmark & 93.02\tiny{$\pm$0.03} & 84.85\tiny{$\pm$0.10}\\
    % \rowcolor{Gray} SCR & SAGN & No &  $\pm$ & $\pm$ \\
    % \rowcolor{Gray} MCR & SAGN & No & 93.25 $\pm$ 0.04 & 84.41 $\pm$ 0.05 \\
    RLU & GAMLP & - & 93.24\tiny{$\pm$0.05} & 84.59\tiny{$\pm$0.10} \\
    % \rowcolor{Gray} \model & MLP & - & 76.51 $\pm$ 0.08 & 62.83 $\pm$ 0.14\\
    % \rowcolor{Gray} \model & SIGN & - & 92.96 $\pm$ 0.08 & 81.64 $\pm$ 0.07\\
    \rowcolor{Gray} \model & GAMLP & - & 93.30\tiny{$\pm$0.06} & 84.07\tiny{$\pm$0.06} \\
    \rowcolor{Gray} \model-m & GAMLP & - & 93.19\tiny{$\pm$0.03} & 84.62\tiny{$\pm$0.03} \\
    \rowcolor{Gray} RLU+\model & GAMLP & - & 92.92\tiny{$\pm$0.05} & 85.05\tiny{$\pm$0.09} \\
    \rowcolor{Gray} RLU+\model & GAMLP & \Checkmark & 93.04\tiny{$\pm$0.05} & \textbf{85.20}\tiny{$\pm$0.08} \\
    \midrule
    \multicolumn{5}{l}{\textit{using node features generated by GIANT-XRT}} \\
    SLE & SAGN & - & 93.63\tiny{$\pm$0.05} & 86.22\tiny{$\pm$0.22} \\
    SLE & SAGN & \Checkmark & 93.52\tiny{$\pm$0.05} & 86.43\tiny{$\pm$0.20} \\
    \rowcolor{Gray} \model & SAGN & - & 93.64\tiny{$\pm$0.05} & 86.67\tiny{$\pm$0.09} \\
    \rowcolor{Gray} \model{} & SAGN & \Checkmark & 93.57\tiny{$\pm$0.04} & \textbf{86.80}\tiny{$\pm$0.07} \\
    \rowcolor{Gray} \model-m & SAGN & - & 93.89\tiny{$\pm$0.02} & 86.51\tiny{$\pm$0.09} \\
    \rowcolor{Gray} \model{}-m & SAGN & \Checkmark & 93.87\tiny{$\pm$0.02} & \textbf{86.73}\tiny{$\pm$0.08} \\
    \bottomrule
  \end{tabular}
\end{table}

\vpara{Baselines.}
\textit{ogbn-products} is a co-purchasing network of Amazon products~\cite{Bhatia16}, and the task is to predict the category of a product.
For comparison, we include classic GNN models and several top-performing methods, including MLP, GCN~\cite{kipf2016semi}, GraphSAGE~\cite{hamilton2017inductive}, SIGN~\cite{frasca2020sign}, SAGN~\cite{sun2021scalable} and GAMLP~\cite{zhang2021graph} as baselines.
We evaluate \model{} with the two GNN models---GAMLP~\cite{zhang2021graph} and SAGN~\cite{sun2021scalable}.\footnote{We also tested the other GNN models. Check Section~\ref{sec:other_gnn} for more details. We choose GAMLP and SAGN as base models due to their top performance on the OGB datasets.}
We also make comparisons of C\&S~\cite{huang2021combining}, which utilizes two post-processing steps, an ``error correlation'' and a ``prediction correlation'', that exploit correlation in the label structure.
C\&S is considered as a standard module for performance improvement especially on the ogbn-product dataset.
In addition, we report the results using the GIANT-XRT~\cite{chien2021node} features that are generated from an encoder trained on an eXtreme Multi-label Classification (XMC) formalism~\cite{shen2020extreme,yu2020pecos,chang2020taming} in the graph domain.

\textit{ogbn-mag} is a heterogeneous academic graph from a subset of the MAG~\cite{wang2020microsoft}. 
%It consists of four types of entities, including papers, authors, institutions, and fields of study, as well as their relations connecting two types of entities. Each paper comes with a 128-dimensional word2vec feature vector obtained by averaging the embeddings of words in its title and abstract, while other types of entities are not associated with node features. 
The task of ogbn-mag is to predict the venue information (conference or journal) of each paper, which is consistent with practical interests of MAG. 
For ogbn-mag, we selected MLP, R-GCN~\cite{schlichtkrull2018modeling}, SIGN~\cite{frasca2020sign}, NARS~\cite{yu2020scalable}, NARS\_SAGN~\cite{sun2021scalable}, and NARS\_GAMLP~\cite{zhang2021graph} as baselines.

% \textit{ogbn-papers100M} is an extremely large citation network with more than 100 million nodes and 1 billion edges.
To demonstrate the scalability of our proposed framework, we conduct an experiment on an extremely large citation network, \textit{ogbn-papers100M}, which has more than 100 million nodes and 1 billion edges.
%Each node is a paper and each directed edge indicates that one paper cites another one. Similar to ogbn-mag, each paper comes with a 128-dimensional feature vector. Approximately 1.5 million papers are arXiv papers with arXiv subject areas as node labels. 
As the ogbn-papers100M dataset is such a giant graph, only a few scalable GNNs could handle it. 
Here, we report the results of MLP, SGC~\cite{wu2019simplifying}, SIGN~\cite{frasca2020sign}, SAGN~\cite{sun2021scalable}, GAMLP~\cite{zhang2021graph}.

The results of these aforementioned baselines are directly taken from the OGB leaderboards.

\vpara{Results.}
Table~\ref{tab:products} summaries the validation and test accuracy of baselines and our methods on ogbn-products.
The prediction accuracy of test set is improved by 0.53\% and 1.08\% by using \model{} and \model{}-m on GAMLP, respectively.
We can also observe that \model{}-m is competitive with RLU, a multi-stage self-training methods. 
In addition, \model{} is more compatible with RLU. Based on GAMLP (RLU), after applying \model{} as a training technique, the performance could be further improved by 0.46\%, which is currently the best performance on the dataset.
After further applying C\&S as the post-processing step, the performance can be improved by 0.15\%. 
By using the node features provided by GIANT-XRT~\cite{chien2021node}, our method can also bring performance gains. Applying \model{} and \model{}-m to the SAGN model, the performance 
exceeds all existing baselines. Compared with the SAGN (RLU), we improve the test accuracy by 0.45\% and 0.29\%. After adding C\&S as post-processing, we improve the test accuracy by 0.37\% and 0.30\%. 
Overall, our proposed framework brings 0.37\% and 0.46\% improvements on ogbn-products with and without GIANT-XRT features, respectively.

Table~\ref{tab:mag} lists the performances of our methods and baseline methods on the ogbn-mag dataset.
We used NARS\_GAMLP as the base model, which shows superior performance than NARS\_SAGN. 
Compared with NARS\_GAMLP, the test accuracy could be improved by 0.36\% and 0.55\% when applying \model{} and \model{}-m. 
Meanwhile, when adding \model{} to NARS\_GAMLP (RLU), the performance can be still improved by 0.41\%.

Table~\ref{tab:paper} shows the performances of all methods on the ogbn-papers100M dataset. 
For the basic architecture, 
By using \model{} and \model{}-m on GAMLP, the test accuracy could be improved by 0.43\% and 0.45\%. 
Moreover, by applying \model{} to GAMLP (RLU), the final performance exceeds all the compared baselines.

\begin{minipage}{\textwidth}
  \centering
  \footnotesize
  \begin{minipage}[t]{0.47\textwidth}
  \makeatletter\def\@captype{table}
  \caption{\label{tab:mag}Accuracy on ogbn-mag.}
  \renewcommand\tabcolsep{3pt}
    \begin{tabular}{llcc}
    \toprule
    Methods & Arch. & Validation & Test \\
    \midrule
    - & \scriptsize{MLP} & 26.26\tiny{$\pm$0.16} & 26.92\tiny{$\pm$0.26} \\
    - & \scriptsize{R-GCN} & 40.84\tiny{$\pm$}0.41 & 39.77\tiny{$\pm$0.46} \\
    - & \scriptsize{SIGN} & 40.68\tiny{$\pm$0.10} & 40.46\tiny{$\pm$0.12} \\
    - & \scriptsize{NARS} & 53.72\tiny{$\pm$0.09} & 52.40\tiny{$\pm$0.16} \\
    - & \scriptsize{NARS\_SAGN} & 54.12\tiny{$\pm$0.15} & 52.32\tiny{$\pm$0.25} \\
    - & \scriptsize{NARS\_GAMLP} & 55.48\tiny{$\pm$0.08} & 53.96\tiny{$\pm$0.18} \\
    \midrule
    SLE & \scriptsize{NARS\_SAGN} & 55.91\tiny{$\pm$0.17} & 54.40\tiny{$\pm$0.15} \\
    RLU & \scriptsize{NARS\_GAMLP} & 57.02\tiny{$\pm$0.41} & 55.90\tiny{$\pm$0.27} \\
    \rowcolor{Gray} \model & \scriptsize{NARS\_GAMLP} & 56.54\tiny{$\pm$0.21} & 54.32\tiny{$\pm$0.18}\\
    \rowcolor{Gray} \model-m & \scriptsize{NARS\_GAMLP} & 55.90\tiny{$\pm$0.28} & 54.51\tiny{$\pm$0.19} \\
    \rowcolor{Gray} RLU+\model & \scriptsize{NARS\_GAMLP} & 57.34\tiny{$\pm$0.35} & \textbf{56.31}\tiny{$\pm$0.21} \\
    \bottomrule
  \end{tabular}
  \end{minipage}
  \begin{minipage}[t]{0.47\textwidth}
  \makeatletter\def\@captype{table}
   \caption{\label{tab:paper}Accuracy on ogbn-papers100M.}
  \renewcommand\tabcolsep{3pt}
  \begin{tabular}{llcc}
    \toprule
    Methods & Arch. & Validation & Test \\
    \midrule
    - & \scriptsize{MLP} & 49.60\tiny{$\pm$0.29} & 47.24 \tiny{$\pm$ 0.31} \\
    - & \scriptsize{SGC} & 66.48\tiny{$\pm$0.20} & 63.29 \tiny{$\pm$ 0.19} \\
    - & \scriptsize{SIGN} & 69.32\tiny{$\pm$0.06} & 65.68 \tiny{$\pm$ 0.06} \\
    - & \scriptsize{SIGN-XL} & 70.32\tiny{$\pm$0.11} & 67.06 \tiny{$\pm$ 0.17} \\
    - & \scriptsize{SAGN} & 70.34\tiny{$\pm$0.99} & 66.75 \tiny{$\pm$ 0.84} \\
    - & \scriptsize{GAMLP} & 71.17\tiny{$\pm$0.14} & 67.71 \tiny{$\pm$ 0.20} \\
    % - & FSGNN & 71.75 $\pm$ 0.07 & 68.07 $\pm$ 0.06 \\
    \midrule
    SLE & \scriptsize{SAGN} & 71.63\tiny{$\pm$0.07} & 68.30\tiny{$\pm$0.08} \\
    RLU & \scriptsize{GAMLP} & 71.59\tiny{$\pm$0.05} & 68.25\tiny{$\pm$0.11} \\
    \rowcolor{Gray} \model & \scriptsize{GAMLP}\;\;\;\;\;\;\;\; &  71.90\tiny{$\pm$0.07} & 68.14\tiny{$\pm$0.08} \\
    \rowcolor{Gray} \model-m & \scriptsize{GAMLP} & 71.86\tiny{$\pm$0.08} & 68.16\tiny{$\pm$0.12} \\
    \rowcolor{Gray} RLU + \model & \scriptsize{GAMLP} & 71.88\tiny{$\pm$0.07} & \textbf{68.42}\tiny{$\pm$0.15} \\
    \bottomrule
  \end{tabular}
  \end{minipage}
\end{minipage}

\subsection{Experiments on Different Graph Neural Networks}
\label{sec:other_gnn}

\begin{wraptable}{r}{8cm}
% \begin{table}[t]
  \caption{Classification accuracy with different GNNs as the base encoder on the ogbn-products dataset. }
  \label{tab:products_across_arch}
    % \footnotesize
    \centering
    \renewcommand\tabcolsep{3pt}
  \begin{tabular}{llll@{\;}}
    \toprule
    Methods & Arch. & Validation & Test \\
    \midrule
    - & MLP & 75.54\tiny{$\pm$0.14} & 61.06 {\tiny$\pm$0.08} \\
    SCR & MLP & 76.51\tiny{$\pm$0.08} & 62.83 {\tiny $\pm$0.14} (+1.77) \\
    \midrule
    - & GraphSAGE & 92.14\tiny{$\pm$0.09} & 78.75 {\tiny $\pm$0.38} \\
    SCR & GraphSAGE & 92.30\tiny{$\pm$0.05} & 79.82 {\tiny $\pm$0.39} (+1.07) \\
    \midrule
    - & ClusterGCN & 91.66\tiny{$\pm$0.06} & 78.37 {\tiny $\pm$0.49} \\
    SCR & ClusterGCN & 91.86\tiny{$\pm$0.08} & 78.72 {\tiny $\pm$0.23} (+0.35) \\
    \midrule
    - & GraphSAINT & 91.67\tiny{$\pm$0.07} & 79.17 {\tiny $\pm$0.30} \\
    SCR & GraphSAINT & 91.71\tiny{$\pm$0.06} & 80.01 {\tiny $\pm$0.33} (+0.84) \\
    % \midrule
    % - & GAT & 91.66 $\pm$ 0.06 & 78.37 $\pm$ 0.49 \\
    % SCR & GAT & 91.86 $\pm$ 0.08 & 78.72 $\pm$ 0.23 \\
    \midrule
    - & SIGN & 92.99\tiny{$\pm$0.04} & 80.52 {\tiny $\pm$0.16} \\
    SCR & SIGN & 92.96\tiny{$\pm$0.08} & 81.64 {\tiny $\pm$0.07} (+1.12) \\
    \bottomrule
  \end{tabular}
% \end{table}
\end{wraptable}

By design, the \model{} framework is a general consistency regularization framework with the goal to improve the base graph models. 
As shown above, it can help both GAMLP and SAGN---the previous top entries on three OGB leaderboards---to improve their performance. 
Herein we examine whether \model{} can broadly help other graph encoders as promised. 
Specifically, we consider the following five architectures, namely MLP, GraphSAGE~\cite{hamilton2017inductive}, ClusterGCN~\cite{ChiangLSLBH19}, GraphSAINT~\cite{zeng2019graphsaint} and SIGN~\cite{yu2020scalable}, as the base graph encoder in \model{}.
Table~\ref{tab:products_across_arch} reports the results of the original models as well as those with \model{} on ogbn-products. 
We observe consistent performance gains brought by the \model{} framework for all five encoders, further demonstrating that \model{} is a general consistency regularization technique that can benefit the performance of graph models. 
%the performance gain brought by the \model framework on the ogbn-products dataset.
%We see a consistent improvement ranging from 0.35\% to 1.77\% in terms of the test accuracy.

\subsection{Comparison of Training Efficiency}

% \begin{wrapfigure}{r}{8cm}
\begin{figure}[t]
  \centering
  \begin{minipage}[t]{0.4\textwidth}
    \includegraphics[width=\textwidth]{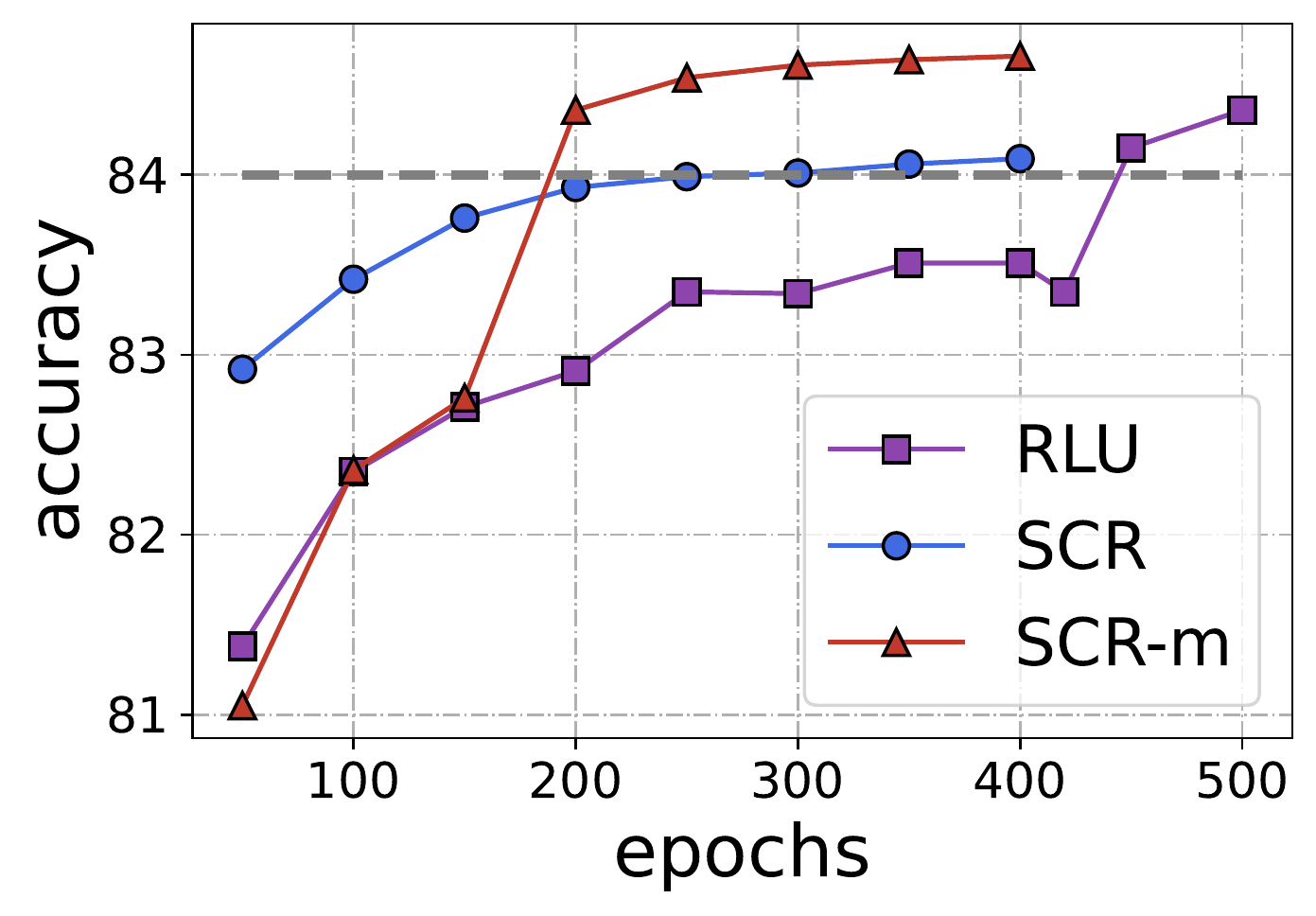}
    \subcaption{ogbn-products}
  \end{minipage}
  \begin{minipage}[t]{0.4\textwidth}
    \includegraphics[width=\textwidth]{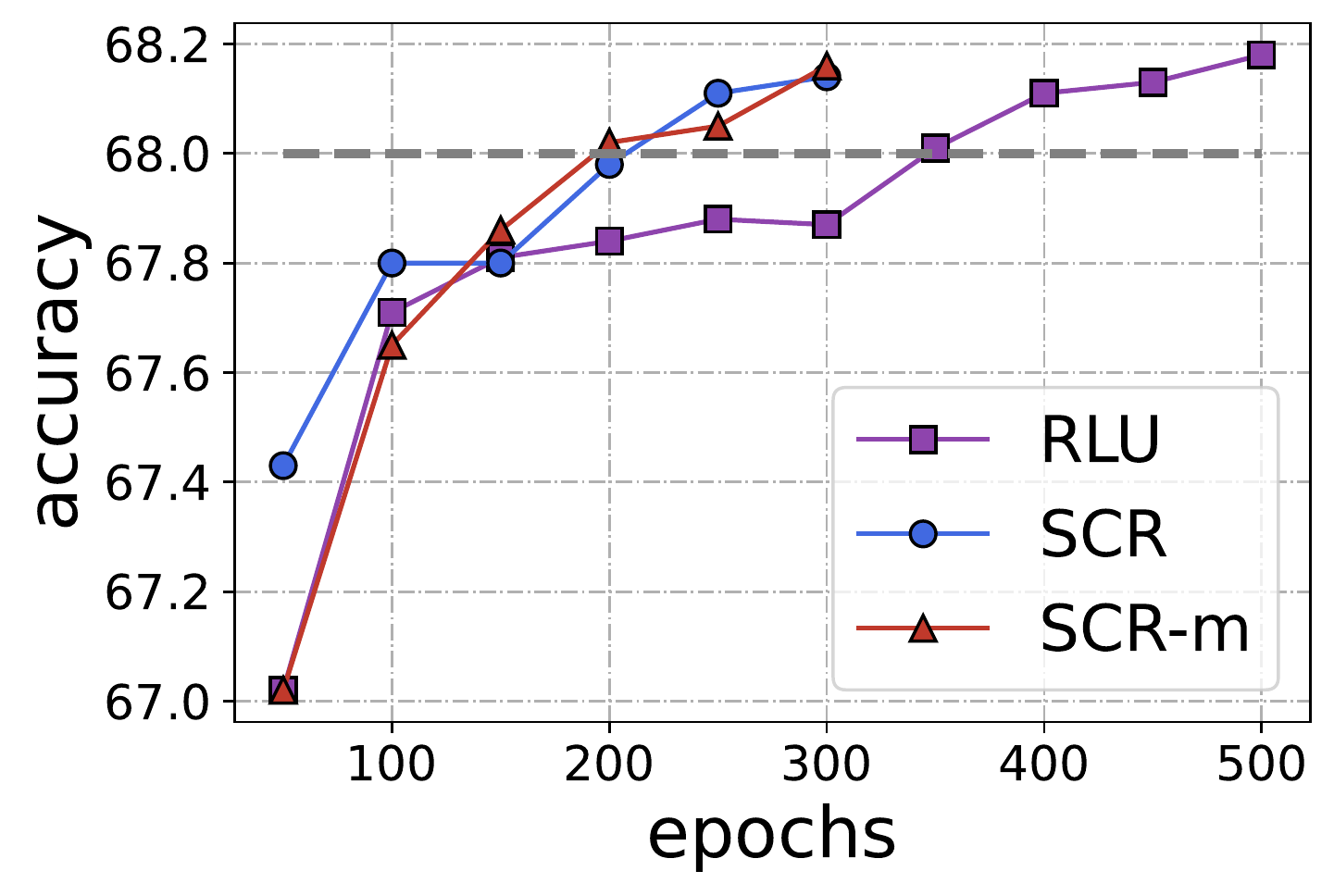}
    \subcaption{ogbn-papers100M}
  \end{minipage}
    % \begin{subfigure}{0.32\textwidth}
    %   \includegraphics[width=\textwidth]{figures/mag.png}
    %   \caption{ogbn-mag}
    % \end{subfigure}
  \caption{Classification accuracy of the \model{} framework and our baseline RLU when training epochs increase.}
  \label{fig:effi}
\end{figure}
% \end{wrapfigure}

Multi-stage training methods, such as RLU, rely on repeated retraining to produce new pseudo labels. Our proposed \model{} and \model{}-m avoid this unnecessary overhead and are therefore more efficient, with a significantly reduced number of training epochs required. To verify the efficiency of our methods, we used GAMLP as the base model and conducted comparative experiments with RLU on three OGB datasets.
Figure~\ref{fig:effi} shows the test accuracy per epoch on the ogbn-products and ogbn-papers100M datasets.
On the ogbn-products dataset, it takes 200 epochs for the \model{}-m to reach the accuracy of 84\% while RLU needs about 450 epochs to reach this level.
On the ogbn-papers100M dataset, it takes around 200 epochs for the \model{} and the \model{}-m to reach the accuracy of 68\% while RLU needs about 350 epochs. 
% \yukuo{The following statement is out-of-date}
% In Figure~\ref{fig:effi}, it can be seen that compared with RLU, training epochs reduces by 69.2\%, 52.9\%, and 45.4\% respectively in the three OGB datasets. The experimental performance can be further improved by adding RLU to SCR. SCR+RLU also outperformed RLU alone in roughly the same training epochs.

% \begin{wrapfigure}[r]{8cm}
% \begin{figure}[t]
%     \centering
%     \begin{subfigure}{0.3\textwidth}
%       \includegraphics[width=\textwidth]{figures/products-v2.pdf}
%       \caption{ogbn-products}
%     \end{subfigure}
%     \begin{subfigure}{0.3\textwidth}
%       \includegraphics[width=\textwidth]{figures/papers100m-v2.pdf}
%       \caption{ogbn-papers100M}
%     \end{subfigure}
%     % \begin{subfigure}{0.32\textwidth}
%     %   \includegraphics[width=\textwidth]{figures/mag.png}
%     %   \caption{ogbn-mag}
%     % \end{subfigure}
%     %\vspace{-2mm}
%     \caption{Classification accuracy of the \model  framework and our baseline RLU when training epochs increase.}
%     %\vspace{-2mm}
%     \label{fig:effi}
% \end{figure}
% \end{wrapfigure}

\subsection{Hyper-parameter Sensitivity}
\begin{figure*}[t]
    \centering
    \subfloat[GAMLP+\model{}]{\label{fig:consis_weight_1}\includegraphics[width=0.25\textwidth]{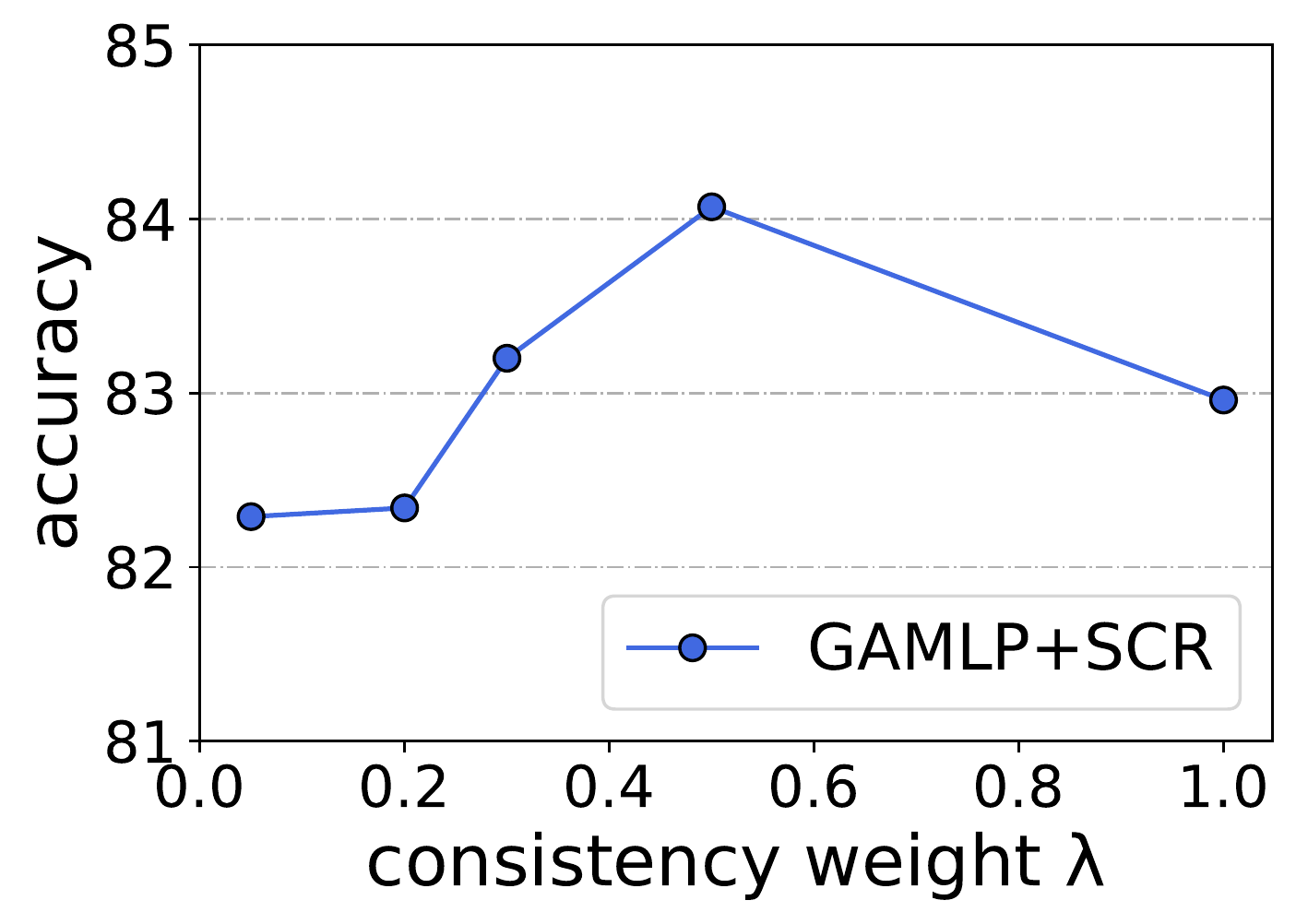}}
    \subfloat[GAMLP + \model{}]{\label{fig:thres}\includegraphics[width=0.25\textwidth]{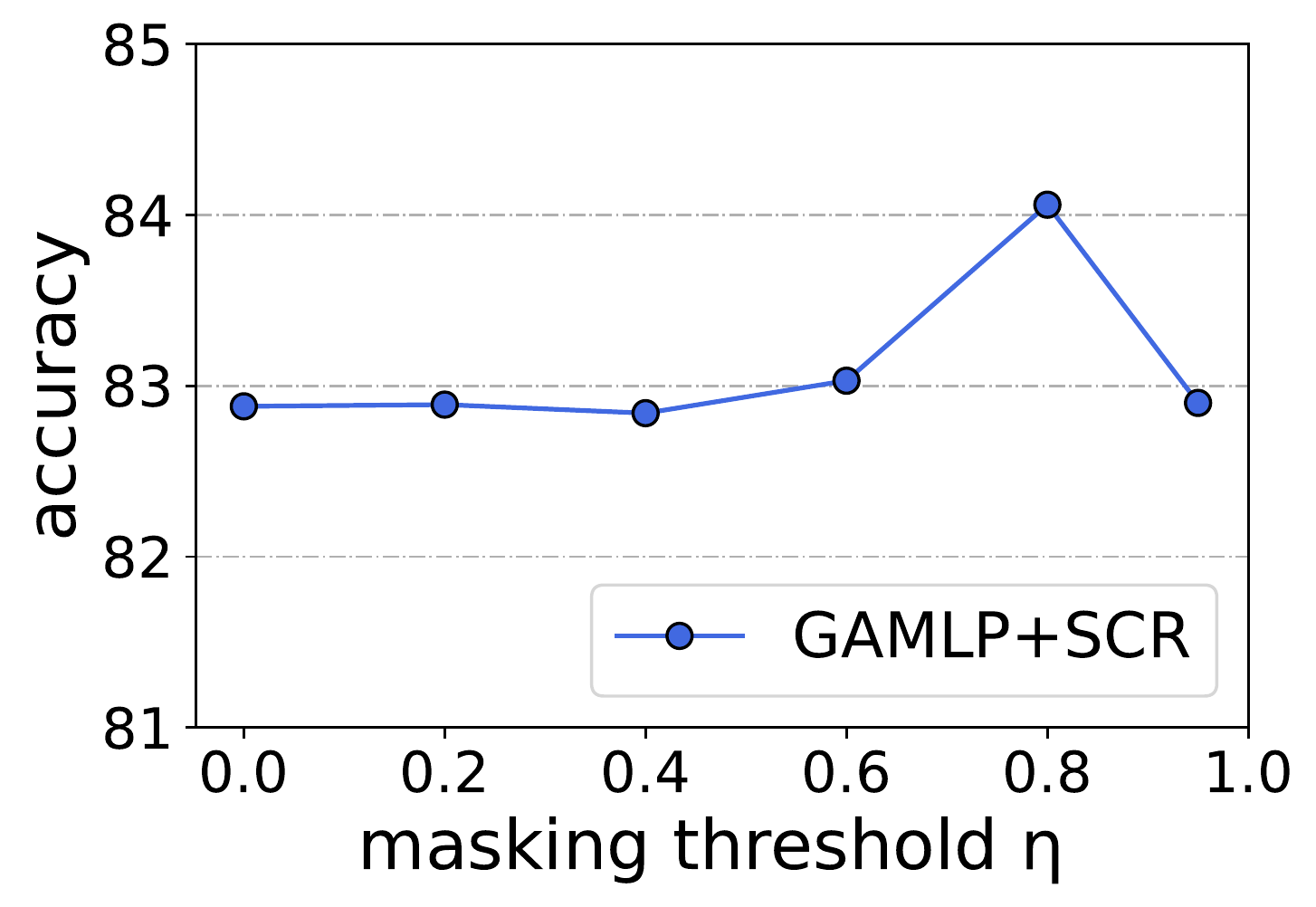}}
    % \caption{Effects of consistency weight $\lambda$ and masking threshold $\eta$ on GAMLP+\model-1}
    \subfloat[GAMLP + \model{}-m]{\label{fig:consis_weight_m}\includegraphics[width=0.25\textwidth]{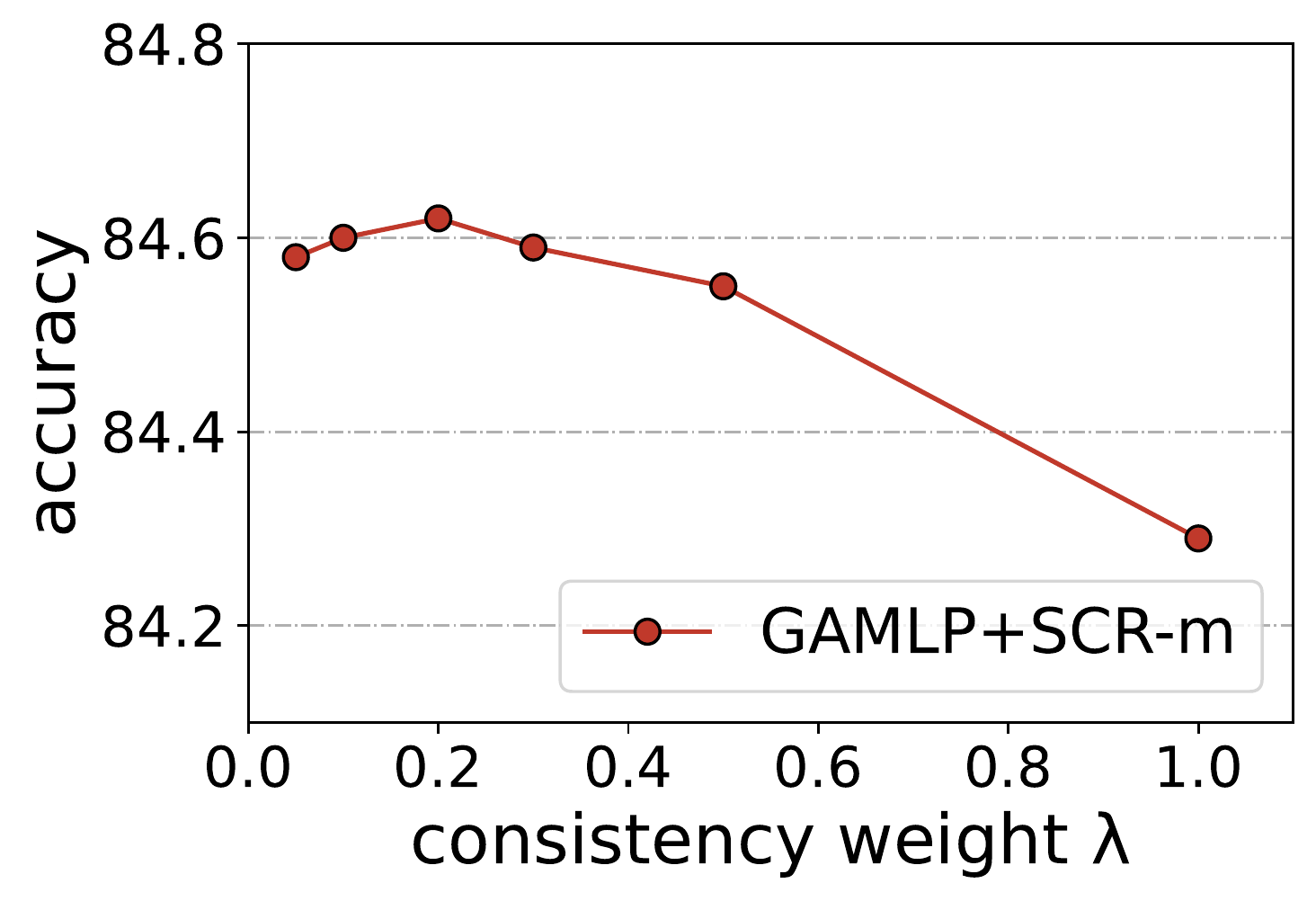}}
    \subfloat[GAMLP + \model{}-m]{\label{fig:ema_decay}\includegraphics[width=0.25\textwidth]{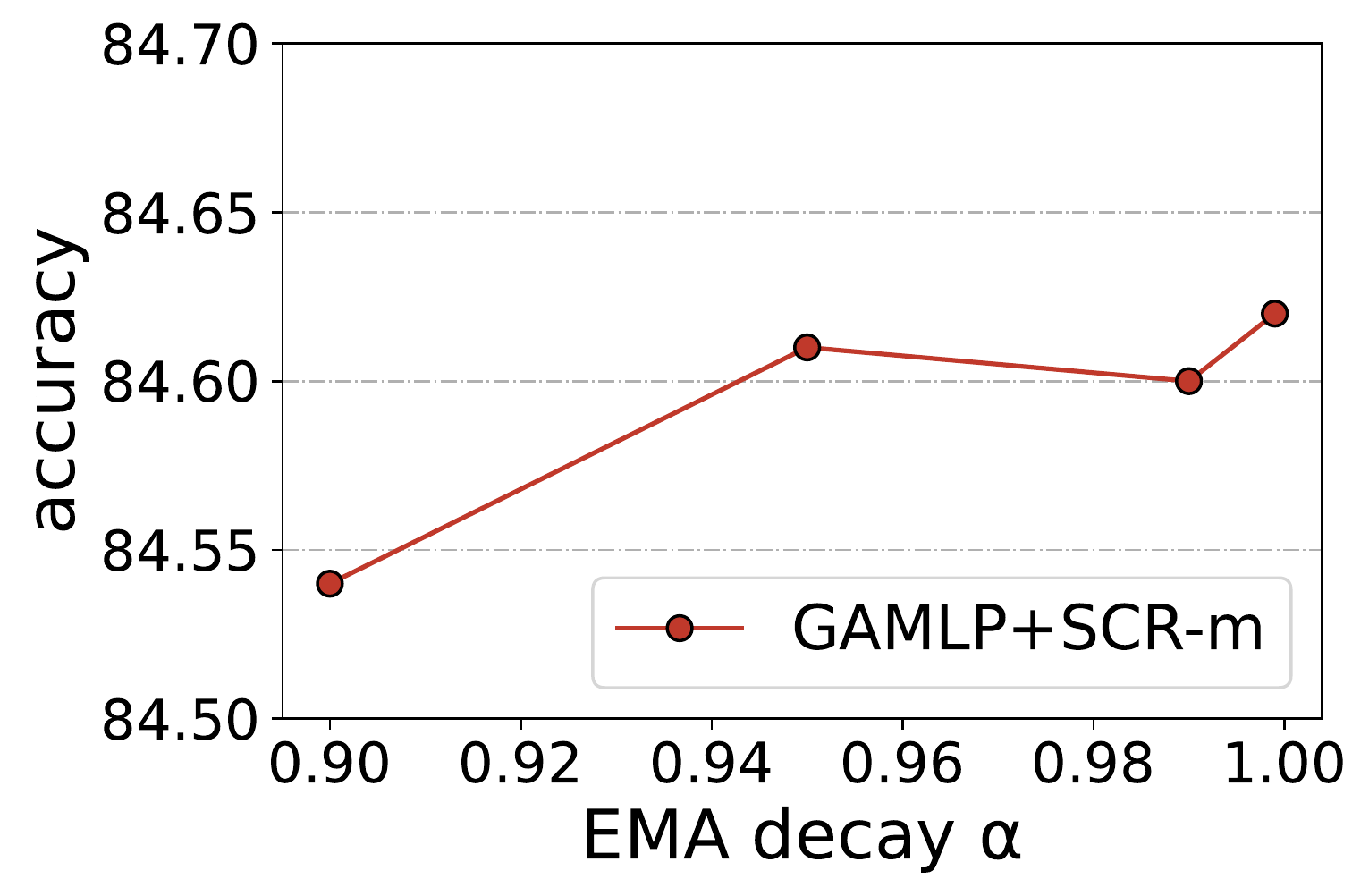}}
    %\vspace{-2mm}
    \caption{Effects of hyper-parameters on GAMLP + \model{} and GAMLP + \model{}-m.}
\end{figure*}

We mainly performed hyper-parameter studies on ogbn-products with 
GAMLP + \model{} and GAMLP + \model{}-m.
In Figure~\ref{fig:consis_weight_1} and Figure~\ref{fig:consis_weight_m}, we analyzed the influence of $\lambda$ of consistency loss on training. It is an important issue to balance the proportion of supervised loss and consistency loss in the process of tuning models. From the results, we can see that too large $\lambda$ or too small $\lambda$ will affect the final performance.
In Figure~\ref{fig:thres}, We explored the influence of masking threshold on experimental results. Too large threshold will reduce the number of nodes involved in calculating consistency loss, and too small threshold will make some bad predictions affect the overall training direction. 
In Figure~\ref{fig:ema_decay}, we analyzed the influence of different EMA(Exponential Moving Average) decay on experimental results. Different decay can be regarded as an ensemble of different numbers of past model parameters. Experimental results show that large decay can avoid the vibration of the model and bring better results. 

%The effects of other hyper-parameters could be found in the Appendix.

% \dong{to chenhui, figure ref?}

%% file: 5.conclusion.tex
\section{Conclusion}%
\label{sec:conclusion}

This paper demonstrates how consistency regularization improves the training performance of graph neural networks under semi-supervised setting. 
% This work also explains that pseudo labeling is a special form of consistency regularization. 
We design \model{}---a simple graph learning consistency regularization technique---as well as its variant \model{}-m. 
With these methods of consistency regularization, the results of state-of-the-art GNN models can be consistently improved. 
At the time of writing, we achieve the top-1 performance on the three large-scale OGB datasets: ogbn-products, ogbn-mag, and ogbn-papers100M. 
Moreover, \model{} (and \model{}-m) are a general regularization framework that by design can take any graph models as the base encoder and thus improve their performance. 
This attempt provides inspiration for further exploration of consistency regularization on graph neural networks. 
In the future, we plan to design more effective consistency regularization strategies for graph-structured data.

\vpara{Limitations of the work.}
Our work assumes that the labeled data and the unlabeled data are generated from the same distribution. However, This assumption is not always hold in real applications. If the distribution mismatch between labeled data and unlabeled data is identifiable, we would not generate reliable pseudo labels for unlabeled data with the model learned from the labeled data.

%% file: 6.appendix.tex
% TODO: remove 
% just for reference
% \begin{table}[tb]
%   \centering
%   \begin{tabular}{c|c}
%     \toprule
%     Symbol & Meaning \\
%     \midrule
%     $\set{V}$ & node set \\ 
%     $\set{E}$ & edge set \\
%     $i$ & node index \\
%     $L$ & labeled data \\
%     $U$ & unlabeled data \\
%     $\mathcal{L} (\cdot)$ & loss function  \\
%     \bottomrule
%   \end{tabular}
%   \caption{Symbols used in this paper}
%   \label{tab:symbols}
% \end{table}

% \begin{table*}
%   \centering
%   \begin{tabular}{@{}llllll@{}}
%     \toprule
%     Algorithm & Updating Rule & Input Augmentation & Prediction Processing & Unsupervised loss & Confident Threshold \\
%     \midrule
%     SLE~\cite{sun2021scalable} & Copy at predefined steps & None & Hard artificial labels & Cross entropy & Yes \\
%     GRAND~\cite{} & Gradient descent & DropNode & Average + Sharpen & Mean square error & No \\
%     GraphMix~\cite{} & Gradient descent & Mixup & Average + Sharpen + Mixup & Cross entropy & No \\
%     Ours & EMA & None & Sharpen & & Yes \\
%     \bottomrule
%   \end{tabular}
%   \caption{Comparison of semi-supervised learning algorithm that explicitly or implicitly involve a teacher-student architecture.}
% \end{table*}

% \section{APPENDIX}
\section{Appendix}
\label{sec:appendix}

\subsection{Connections with Previous Work}

\vpara{GRAND~\cite{feng2020graph}} can be also viewed as a consistency regularization method for graph data.
It utilizes random propagation, a data augmentation technique designed for graph data, to generate noisy predictions.
In GRAND, every noisy prediction relies on a single run of the random propagation.
To train a neural network with GRAND for $N_t$ training steps, we need to run the random propagation for $N_t \cdot S$ times.
The total complexity of the random propagation is $\mathcal{O} (N_t S K d (N + M))$, where $K$, $d$, $N$, $M$ are the numbers of the propagation steps, node features, nodes and edges, respectively. 
%This complexity is high, in particular when the graph is large.
In practice, it is computationally infeasible to perform this operation for large graphs. 
In addition, the random propagation in GRAND needs to load the entire graph into memory at each training step, making its scalability constrained by the memory size. 
%to million-scale or larger networks.
In contrast, 
the noisy predictions in \model{} do not rely on the randomness of the propagation. 
Therefore, we only need to run the propagation 
%for only 
one time and the propagated node features can be reused in the following training steps, which results in a complexity of $\mathcal{O} (K d (N + M))$, which is practically much cheap. 
To conclude, the performance of \model{} does not rely on the expensive data augmentation procedure used in GRAND, enabling it to scale to any size of networks as long as the underlying graph encoder can handle them. 
%framework simplifies GRAND by removing the data augmentation module.

\vpara{R-Drop~\cite{Liang21drop}} is a regularization method to reduce the inconsistency between the training and inference stages caused by dropout.
The key idea is to minimize the Kullback-Leibler divergence between two outputs generated by the same input data.
The R-Drop regularization is applied in the supervised setting where each data point has a ground-truth target.
Thus, the pseudo labeling is unnecessary for R-Drop.
In the \model{} framework, the consistency regularization is applied to the unlabeled data, and uses pseudo labels to align model outputs, making it fundamentally different from R-Drop. 

\vpara{SLE~\cite{sun2021scalable} \& RLU~\cite{zhang2021graph}} utilize unlabeled nodes by self training.
When applying SLE or RLU to train a GNN model, the training phase is divided into several stages.
In the first stage, only the labeled data is used to train the model.
From the second stage, the model trained at the previous stage is used to label nodes that have not been labeled yet.
The training data is enlarged by adding node-label pairs with high prediction confidence.
A new model will be trained on the enlarged training data.
As it needs to train a model to converge at each stage, the training time is much longer than \model{}.

\subsection{Implementation Notes}
\label{sec:impl}

\vpara{Running Environment.} 
Our proposed framework is implemented via PyTorch~\cite{paszke2019pytorch} and CogDL~\cite{cen2021cogdl}, a toolkit for deep learning on graphs. For our methods, the experiments are conducted on a linux machine with Intel(R) Xeon(R) Gold 6240 CPU @ 2.60GHz, 377G RAM, and 10 NVIDIA GeForce RTX 3090 with 24GB GPU memory. 
For a fair comparison, we repeat each experiment ten times and report the mean and standard deviation except for the largest ogbn-papers100M dataset. For the ogbn-papers100M dataset, we run each method three times following the official requirements of OGB leaderboards. As for software versions, we use Python 3.8, PyTorch 1.7.1, CogDL 0.5.2, OGB 1.3.2, and CUDA 11.2.

\vpara{Implementation Details.}
% As suggested by OGB, we repeat each experiment for 10 times with random seeds and report the average and the standard deviation of the classification accuracy. 
% On ogbn-papers100M, the reported results are based on 3 independent runs. 
% The results of baselines are directly taken from the OGB leaderboards.
We adopt the \model{} framework on 7 GNN architectures.
The codes are adapted from the following repositories:
\begin{itemize}
  \item MLP: \url{https://github.com/snap-stanford/ogb/blob/master/examples/nodeproppred/products/mlp.py}
  \item GraphSAGE: \url{https://github.com/pyg-team/pytorch_geometric/blob/master/examples/ogbn_products_sage.py}
  \item ClusterGCN: \url{https://github.com/snap-stanford/ogb/blob/master/examples/nodeproppred/products/cluster_gcn.py}
  \item GraphSAINT: \url{https://github.com/snap-stanford/ogb/blob/master/examples/nodeproppred/products/graph_saint.py}
  \item SIGN: \url{https://github.com/dmlc/dgl/blob/master/examples/pytorch/ogb/sign/sign.py}
  \item SAGN: \url{https://github.com/skepsun/SAGN_with_SLE}
  \item GAMLP: \url{https://github.com/PKU-DAIR/GAMLP}
\end{itemize}

% \vpara{GNN Architectures.}
% We mainly utilize two SOTA architectures for our experiments, that is, SAGN and GAMLP.
% We use the code from their official repositories. 
% \begin{itemize}
%     \item SAGN: \url{https://github.com/skepsun/SAGN_with_SLE}
%     \item GAMLP: \url{https://github.com/PKU-DAIR/GAMLP}
% \end{itemize}

% \subsection{Detailed Hyperparameters}

\subsection{Hyperparameter Settings}
\label{sec:hparam}

For all experiments, we set the hyperparameters related to the model architecture, such as hidden size, model depth, and propagation steps, as the defaults.
The hyperparameters introduced by the \model{} framework are listed as follows:
\begin{itemize}
  \item $\lambda$: the weight of the consistency regularizer
  \item $\mathtt{dist}$: distance function
  \item $S$: the number of views
  \item $\eta$: threshold for confidence-based masking
  \item $T$: the sharpening temperature
  \item $\beta$: update frequency for confidence-based masking
  \item $\alpha$: EMA decay rate, only applied on \model{}-m
  \item $\tau$: warmup epochs, only applied on \model{}-m
\end{itemize}
We set $T$ as $0.5$ and $\beta$ as 20 across all experiments, as the \model{} framework is not very sensitive to them in the experiments.

Table~\ref{tab:hparam_scr} lists the hyperparameters corresponding to the results on the \model{} framework reported in Table~\ref{tab:products}, \ref{tab:mag} and \ref{tab:paper}.
The confidence threshold $\eta$ in \model{}-m is linearly reduced from the initial value to the minimum value.
\begin{table}[thb]
  \centering
  \caption{\label{tab:hparam_scr}Detailed hyperparameter setting on \model{}.}
  \begin{tabular}{@{}c@{\:}c@{\:}c@{\:}c@{\hspace{0.8em}}c@{\:}c@{\:}c@{}}
    \toprule
    & \multicolumn{3}{c}{SCR} & \multicolumn{3}{c}{SCR-m} \\
    \cmidrule(r{0.8em}){2-4} \cmidrule{5-7}
    Datasets & ogbn-products & ogbn-mag & ogbn-papers100M & ogbn-products & ogbn-mag & ogbn-papers100M \\
    \midrule
    $\lambda$ & 0.5 & 0.1 & 0.2 & 0.2 & 0.025 & 0.03 \\
    $\mathtt{dist}$ & MSE & MSE & MSE & KL & KL & KL \\
    $S$ & 2 & 2 & 2 & 1 & 1 & 1 \\
    $\eta$ & 0.85 & 0.5 & 0.9 & 0.9-0.8 & 0.5-0.4 & 0.9-0.8 \\
    $\alpha$ & --- & --- & --- & 0.999 & 0.99 & 0.99 \\
    $\tau$ & --- & --- & --- & 150 & 60 & 100 \\
% $T$ & 0.5 & 0.5 & 0.5\\
% epoch & 400 & 400 & 300 \\
% hidden size & 512 & 512 &1280 \\
% batch size  & 50000 & 10000 & 5000\\
    \bottomrule
  \end{tabular}
\end{table}

% We provide the detailed hyperparameter settings of \model, RLU+\model and \model-m in Table~\ref{tab:hyper on scr}, Table~\ref{tab:hyper on scrrlu}, and Table~\ref{tab:hyper on mcr}, respectively. 
% Table~\ref{tab:hyper on mcr with giant} shows the configurations of \model and \model-m with GIANT-XRT features. 
% Except for the experiment using node features generated by GIANT-XRT, which used SAGN as the base model, the other experiments used GAMLP as the base model. Other hyperparameters not mentioned are set according to the baseline method. It is especially worth mentioning that the $avgmse$ loss function we use in SCR refers to averaging the distribution of all views. The L2 distance between sharpened label distribution and average distribution is then calculated. Masking threshold $\eta$ used in SCR-m, is a progressively decreasing range. Early stopping patience of 100 epochs was used in most of the experiments.

% Table~\ref{tab:hyper on mcr} lists the hyperparameters used for the experiments on \model{}-m.
% For these experiments, we use the Kullback-Leibler divergence as the distance function, and set $S = 1$, $T = 0.5$, $\beta = 10$.

\begin{table}[tb]
%\begin{wraptable}{r}{8cm}
  \centering
  \caption{\label{tab:hparam_rlu_scr}Detailed hyperparameter setting on RLU + \model{}.}
  \begin{tabular}{cccc}
    \toprule
    Datasets & ogbn-products & ogbn-mag & ogbn-papers100M\\
    \midrule 
    $\lambda$ & 0.1 & 0.07 & 0.1 \\
    $\eta$ & 0.995 & 0.4 & 0.9 \\
% temp & 0.5 & 0.5 & 0.5\\
% stages & 6 & 7 & 4\\
% epochs & 1900  & 1450  & 1600 \\
% hidden size & 512 & 512 &1280 \\
% batch size  & 50000 & 10000 & 5000\\
    \bottomrule
  \end{tabular}
%\end{wraptable}
\end{table}

Table~\ref{tab:hparam_rlu_scr} lists the hyperparameters used for the experiments on RLU + \model{}.
For these experiments, we use the mean squared error as the distance function, and set $S = 2$.
The training stage is set as follows:
\begin{itemize}
  \item On the ogbn-products dataset, we split the training process into 6 stages.
The first stage has 400 epochs and the rest has 300 epochs for each.
  \item On the ogbn-products dataset, we split the training process into 6 stages.
The first stage has 400 epochs and the rest has 300 epochs for each.
  \item On the ogbn-papers100M dataset, we split the training process into 4 stages and each has 400 epochs.
\end{itemize}

%\begin{figure}[htb]
%    \centering
%    \begin{subfigure}{0.23\textwidth}{%
%      \label{fig:temp_1}
%      \includegraphics[width=\textwidth]{figures/scr1-temp.pdf}
%      \caption{Effect of $T$ on GAMLP+\model}
%      }
%    \end{subfigure}
%    \begin{subfigure}{0.23\textwidth}{%
%      \label{fig:temp_m}
%      \includegraphics[width=\textwidth]{figures/scrm-temp.pdf}
%      \caption{Effect of $T$ on GAMLP+\model-m}
%     }
%    \end{subfigure}
%    \caption{Effects of other hyper-parameters on GAMLP+\model and GAMLP+\model-m.}
%    \label{fig:other_hyper}
%\end{figure}

\begin{comment}
\begin{table}[htb]
\centering
\caption{\label{tab:hyper on mcr}Detailed hyperparameter setting on \model{}-m.}
\begin{tabular}{c|ccc}
\toprule
%\multirow{2}*{Datasets} & \multicolumn{2}{c}{ogbn-products} &\multirow{2}*{ogbn-mag} &\multirow{2}*{ogbn-papers100m}\\
Datasets & ogbn-products & ogbn-mag & ogbn-papers100M  \\
%\cline{2-3}
%\multicolumn{1}{c|}{}&GAMLP&SAGN\\
\midrule 
$\lambda$ & 0.2 & 0.025 & 0.03  \\
% loss type & kl & kl & kl \\
$\eta$ & 0.9-0.8  & 0.5-0.4& 0.9-0.8 \\
% temp $T$ & 0.5 & 0.5 & 0.5\\
% gap $\beta$ & 10 & 10 & 10\\
$\alpha$ & 0.999 & 0.99 & 0.99\\
$\tau$ & 150& 60 & 100\\
% epochs & 400 & 400 & 300\\
% hidden size & 512 & 512 &1280 \\
% batch size  & 100000 & 10000 & 5000\\
\bottomrule
\end{tabular}
\end{table}
\end{comment}

%\begin{table}[tb]
\begin{wraptable}{r}{4.5cm}
  \centering
  \caption{\label{tab:hparam_giant}Detailed hyperparameter setting using GIANT-XRT features.}
  \begin{tabular}{ccc}
    \toprule
    & \model{} & \model{}-m\\
    \midrule 
    $\lambda$& 0.2 & 0.02   \\
% loss type & kl & kl  \\
    $S$ & 2 & 1 \\
    $\eta$ & 0.85-0.8 & 0.85-0.75  \\
% temp & 0.5 & 0.5 \\
% gap $\beta$ & 20 & 20 \\
    $\alpha$ & --- & 0.99 \\
% warmup & 100& 100 \\
% epochs & 400  & 400 \\
% hidden size & 256 & 256 \\
% batch size  & 50000 & 50000 \\
    \bottomrule
  \end{tabular}
\end{wraptable}
%\end{table}

Table~\ref{tab:hparam_giant} lists the hyperparameters used for the experiments using node features generated by GIANT.
For these experiments, we use the Kullback-Leibler divergence as the distance function.